\documentclass[a4paper,11pt]{article}
\pdfoutput=1 % if your are submitting a pdflatex (i.e. if you have
\usepackage{jheppub} % for details on the use of the package, please
\usepackage{tikz}
\usepackage{pgfplots}
\usetikzlibrary{calc,decorations.pathmorphing,shapes}
\usepackage{standalone}
\usepackage[T1]{fontenc} % if needed

\title{\boldmath The action of HRT-areas as operators in semiclassical gravity}

\author{Molly Kaplan,} 
\author{Donald Marolf}
\affiliation{Department of Physics, University of California, Santa Barbara, CA 93106, USA}

\emailAdd{mekaplan@ucsb.edu}
\emailAdd{marolf@ucsb.edu}

\abstract{We study the action of Hubeny-Rangamani-Takayanagi (HRT) area operators on the covariant phase space of classical solutions. It has been previously proposed that this action generates a transformation which, roughly speaking, boosts the entanglement wedge on one side of the HRT surface relative to the entanglement wedge on the other side.  We give a sharp argument for a precise result of this form in a general theory of Einstein-Hilbert gravity minimally coupled to matter, taking appropriate care with asymptotically Anti-de Sitter (AdS) boundary conditions. The result agrees with direct computations of commutators involving HRT areas in pure 2+1 dimensional Einstein-Hilbert gravity on spacetimes asymptotic to planar AdS. We also clarify the sense in which this transformation is singular in the deep UV when the HRT-surface is anchored to an asymptotically AdS boundary.
}

\arxivnumber{2203.04270}

\begin{document}
\maketitle
\flushbottom

%%%%%%%%%%%%%%%%%%%%%%%%%%%%%%%%%%%%%%%%%%%%%%%%%%%%%%%%%%%%%%%%%%%%%%%%%%%%%%%%%%%
\section{Introduction}

A fundamental aspect of gauge-gravity duality is the relation between gauge theory entropies and the areas of codimension-2 bulk extremal surfaces described by the Ryu-Takayanagi (RT) correspondence \cite{Ryu:2006bv,Ryu:2006ef} and its covariant Hubeny-Rangamani-Takayanagi (HRT) generalization \cite{Hubeny:2007xt}.  The quantity $A_{HRT}[R]$ defined by computing the area of the HRT surface associated with an appropriate boundary region $R$ may thus be expected to be of great interest in the bulk theory, even without reference to the gauge theory dual.

At the classical level in the bulk,  we can think of this
$A_{HRT}[R]$  as a function on the space of solutions or, equivalently, on either the canonical or covariant  phase space.  At the quantum level, it should define a corresponding quantum operator.   The purpose of this work is to better explore the commutation relations of such operators, either with themselves or with other objects of interest.  We will work at leading order in the bulk semiclassical approximation, where such commutators are described by Poisson brackets, or equivalently by Peierls brackets \cite{Peierls:1952cb} up to the usual factor of $i$.

There is in fact a lengthy history of suggestions that taking brackets with $A_{HRT}[R]$ should generate a transformation closely related to the boost symmetry of a  Rindler wedge in Minkowski space.  Indeed, long before the days of gauge/gravity duality it was noted in various contexts that the area of black hole horizons seemed to generate such transformations; see especially \cite{Carlip:1993sa}, but similar observations are implicit in \cite{Thiemann:1992jj,Kastrup:1993br,Kuchar:1994zk}.

Later, in the context of gauge/gravity duality, analogous suggestions for general HRT-areas $A_{HRT}[R]$ were motivated in \cite{Jafferis:2014lza,Ceyhan:2018zfg,Faulkner:2018faa,Bousso:2019dxk,Bousso:2020yxi} by comparison with modular Hamiltonians, as the latter are again known to act as boosts in appropriate circumstances; see in particular \cite{Lewkowycz:2018sgn} and \cite{Chen:2018rgz}.  In many cases this analogy was based on the Jafferis-Lewkowycz-Maldacena-Suh (JLMS) relation explicitly relating bulk areas to modular Hamiltonians in the gauge theory \cite{Jafferis:2015del}.  Furthermore, in a parallel series of developments, various related results  \cite{Donnelly:2016auv,Speranza:2017gxd,Chandrasekaran:2019ewn} were established in contexts where boundary conditions are imposed at finite-distance boundaries.  In particular, when the boundary is an appropriate bifurcate null surface, the area of the bifurcation surface is known to generate a boost-like symmetry of the associated gravitational system.

Nevertheless, despite the long list of closely related results and arguments given above, it appears that a direct analysis of the action of $A_{HRT}[R]$ on the gravitational phase space has yet to be performed.  Here we are explicitly interested in the case where the relevant HRT surface $\gamma_R$  is determined dynamically and lives in the interior of the system, as opposed to being specified by hand to live on a finite-distance boundary.  Our work will fill this gap and then study the implications for simple commutators involving HRT-areas.

In doing so, we will also give proper consideration to the asymptotically AdS boundary conditions that are of primary interest in the RT and HRT correspondences.  In particular,
in the presence of an asymptotically AdS boundary, the area of a codimension-2 surface anchored to the boundary will generally diverge.  In order to discuss finite quantities, in that context we use $A_{HRT}$ below to denote the renormalized HRT-area given by introducing a cutoff $\epsilon$, subtracting an appropriate covariant counterterm from the naive area, and then sending $\epsilon \rightarrow 0$.  Since the counter-term is a $c$-number, this object generates the same Hamiltonian flow as the naive (unrenormalized) HRT-area.  One should also be aware that, as a result of this renormalization, in even boundary dimensions our $A_{HRT}$ will transform anomalously under conformal transformations.
In contrast, when the boundary anchors are the empty set, no renormalization is needed and we use $A_{HRT}$ to denote the naive area of the HRT surface.

We begin in section \ref{sec:kink} with a direct computation of the flow generated by $A_{HRT}[R]$ using the canonical formalism of Einstein-Hilbert gravity with arbitrary minimally coupled matter.  We study the action of this flow on the initial data on a Cauchy slice $\Sigma$ that runs through the HRT surface $\gamma_R$, showing that it leaves the induced metric unchanged and that it shifts one component of the extrinsic curvature by a delta-function at $\gamma_R$.  This result was predicted in \cite{Bousso:2019dxk,Bousso:2020yxi}, where it was argued to correspond to an operation that, in an appropriate sense, boosts the entanglement wedge of $R$ relative to that of the complementary region $\bar R$.  As a result, on such Cauchy surfaces HRT-area flow also agrees in the bulk with the `kink transformation' introduced in \cite{Bousso:2020yxi}, though (as we review) the two act differently in both the past and future of the HRT surface $\gamma_R$.

The above results and relations are then used in section \ref{sec:AdS3} to derive explicit formulae for the action of HRT-area flow on the AdS$_3$ Poincar\'e vacuum, and in particular to study the action on the boundary stress tensor and on other HRT areas evaluated on that solution.  A particular result is that, while an explicit such flow can be defined for any HRT surface $\gamma_R$, the flow turns out to cause the total energy to diverge when $\gamma_R$ has non-trivial anchors on the AdS boundary.  This is a concrete manifestation of the UV issues foreshadowed in \cite{Ceyhan:2018zfg,Faulkner:2018faa,Bousso:2019dxk}.   

For comparison, section \ref{sec:conf} then provides an independent computation of the associated commutators evaluated on general solutions of pure 2+1 Einstein-Hilbert gravity asymptotic to Poincar\'e AdS$_3$.  Instead of using the canonical commutation relations in the bulk, this latter approach is based on the fact that the above solutions can be constructed by acting on the Poincar\'e vacuum with boundary conformal transformations.   From this it follows that any observable can be expressed in terms of the boundary stress tensor, so that the stress tensor algebra can be used to compute general commutators.   We close with some final comments and future directions in section \ref{sec:disc}.

\section{HRT-area flow as a boundary-condition-preserving kink transformation}\label{sec:kink}

We now derive the Hamiltonian flow generated by HRT-area operators by directly computing Poisson/Peierls brackets  in asymptotically AdS$_D$ Einstein-Hilbert gravity.  In the rest of this work we refer to such brackets as ``semiclassical commutators'' despite the lack of a factor of $i = \sqrt{-1}$.  The commutators for which such computations are straightforward will in fact describe the effect of HRT-flow on certain Cauchy data for the solution, whence the action on the full solution is to be determined by solving the equations of motion.  We thus begin by studying the effect on the desired Cauchy data in section \ref{sec:Cauchydataflow}. Section \ref{sec:BCs} then addresses details of the boundary conditions which determine the full solution.  Finally, section \ref{sec:kinkrel} will discuss the relation to the kink transformation of \cite{Bousso:2020yxi}, which will be useful in deriving further explicit results in section \ref{sec:AdS3}.

As usual, we take the HRT surface $\gamma_R$ to be defined by some region $R$ on the asymptotically AdS$_D$ boundary.  In particular, $R$ is an achronal surface on the boundary and $\gamma_R$ is a codimension-2 extremal surface in the bulk that is anchored to the boundary $\partial R$ of $R$.  Since $\gamma_R$ is an HRT surface, it is in fact the smallest such extremal surface satisfying the homology constraint of \cite{Headrick:2007km}.  The area of $\gamma_R$ thus defines a function on the space of solutions that we may call $A_{HRT}[R]$.

Equivalently, we may think of $A_{HRT}[R]$ as a function on the covariant or canonical gravitational phase space.  To maximize accessibility to most readers, we will take the canonical perspective below.  Since our argument in this section is based solely on the canonical commutation relations of Einstein-Hilbert gravity, all results in this section remain valid in the presence of arbitrary minimally-coupled matter fields.

\subsection{HRT-area flow on a Cauchy surface containing $\gamma_R$}
\label{sec:Cauchydataflow}

The object $A_{HRT}[R]$ is of course fully determined by the spacetime metric $g$.
However, in practice it can useful to evaluate $A_{HRT}[R]$ in two steps, first finding the extremal surface $\gamma_R$ and then computing the area of $\gamma_R$.  In reference to this two-step process, we will write $A_{HRT}[R] = A[\gamma_R, g]$.  In particular, in this way we can think of $A_{HRT}[R]$ as a special case of a more general functional $A[\gamma, g]$ which would compute the area of an arbitrary surface $\gamma$, and where $A_{HRT}$ is obtained from $A[\gamma, g]$  by choosing $\gamma = \gamma_R$ as defined by the given metric $g$.  We can make this very explicit by writing
\begin{equation}
A_{HRT}[R] = A[\gamma, g]|_{\gamma = \gamma_R[g]}.
\end{equation}

The fact that $\gamma_R$ is an extremal surface means that, if we fix the spacetime metric $g$ and vary $A[\gamma, g]$ with respect to $\gamma$, the result vanishes when evaluated at $\gamma=\gamma_R$:
\begin{equation}
\label{eq:extreme1}
\frac{\delta A[\gamma, g]}{\delta \gamma} \biggr|_{\gamma = \gamma_R[g]} = 0.
\end{equation}

The relation \eqref{eq:extreme1} will enter in a critical way into our derivation of HRT-area flow below.  The key point that allows it to be useful is that semiclassical commutators are defined by the Poisson Bracket (or equivalently by the Peierls Bracket \cite{Peierls:1952cb}), which satisfies the Leibniz rule
\begin{equation}
\{B,C\} = B_{,I}C_{,J} \{\zeta^I, \zeta^J\},
\end{equation}
where the $\zeta^I$ are any set of coordinates on phase space and where $B_{,I}$ and $C_{,J}$ denote appropriate (perhaps functional) derivatives of $B,C$ with respect to such coordinates.    Setting $B=A_{HRT}[R]$, we may evaluate its $\zeta^I$ derivatives by first separately varying $A[\gamma, g]$ with respect to $\gamma$ and $g$ and then using the chain rule to relate variations of $\gamma$ and $g$ to variations of the $\zeta^I$.  We thus write
\begin{equation}
\label{eq:chain1}
\frac{\delta}{\delta \zeta^I}A_{HRT}[R] = \frac{\delta A[\gamma, g]}{\delta \gamma}\biggr|_{\gamma = \gamma_R[g]} \frac{\delta \gamma_R[g]}{\delta \zeta^I} + \frac{\delta A[\gamma, g]}{\delta g}\biggr|_{\gamma = \gamma_R[g]} \frac{\delta g}{\delta \zeta^I}.
\end{equation}
The notation implies an appropriate summation over the degrees of freedom associated with the surface $\gamma$ and the spacetime metric $g$.  In particular, the last term in \eqref{eq:chain1} includes both a sum over components of  $g$ at each spacetime point and an integral over spacetime points.

Since the first term in \eqref{eq:chain1} vanishes due to \eqref{eq:extreme1}, we are left only with the second.  This is precisely the statement that semiclassical commutators of $A_{HRT}$ can be computed as if the surface $\gamma_R$ were fixed and did not in fact depend on the phase space coordinates $\zeta^I$.  In other words, it suffices to compute commutators with $A[\gamma, g]$ for some fixed $\gamma$ (say, given by certain coordinate conditions) and then to simply set $\gamma=\gamma_R$ at the end of the calculation.  Note that the final result after setting $\gamma=\gamma_R$ will describe a flow generated by a diffeomorphism-invariant observable, and will thus necessarily map solutions to solutions, even if this is not manifest in the intermediate steps. In particular, in the language of the Hamiltonian formalism, the final flow will necessarily preserve all constraints.

Indeed, since $\gamma_R$ is spacelike, in the canonical formalism we are free to simply suppose that we are given a Cauchy surface $\Sigma$ and a fixed submanifold $\gamma \subset \Sigma$.  We may then take our phase space coordinates $\zeta^I$ to be the induced metric $h_{ij}$ on $\Sigma$ and the (undensitized) gravitational momentum $\Pi^{ij}= \frac{1}{16\pi G} (K^{ij} - K h^{ij})$, where $K^{ij}$ is the extrinsic curvature of $\Sigma$ and $K=K^{ij} h_{ij}$.  Such phase space coordinates have the standard Poisson Brackets
\begin{equation}
    \begin{split}
        \{h_{kl}(x),h^{ij}(y)\}=&0, \\
        \{h_{kl}(x),\Pi^{ij}(y)\}=& \frac{1}{\sqrt{h(y)}}\delta_{(k}^i\delta_{l)}^j\delta^{(D-1)}(x-y),
    \end{split}
\end{equation}
where $x^i$ (or equivalently $y^j$) denotes $D-1$ coordinates on $\Sigma$ and we have used the standard Dirac delta function in terms of the coordinates $x,y$.

Since we choose $\gamma \subset \Sigma$, our $A[\gamma,g]$ will be independent of $\Pi^{ij}$ and will depend only on $h_{ij}$.  Thus $A[\gamma, g]$  commutes with $h_{ij}$ at leading order in the semiclassical expansion, and the leading semiclassical commutator of $A[\gamma, g]$ with any function is determined by $\{A[\gamma,g],\Pi^{ij}\}$, or equivalently by $\{A[\gamma,g],K^{ij}\}$.  We shall keep only such leading-order terms below.

Let us consider the bracket with $K^{ij}$, as it will turn out to yield a geometric interpretation of the flow generated by $A_{HRT}[R]$. Using
\begin{equation}
    K^{ij}=16 \pi G \biggr( \Pi^{ij}+\frac{1}{2-D}\Pi h^{ij} \biggr)
\end{equation}
with  $\Pi=\Pi^{ij}h_{ij}$, one finds
\begin{equation}
\label{eq:hKPB}
    \begin{split}
        \{h_{kl}(x),K^{ij}(y)\} =& \frac{16\pi G}{\sqrt{h(y)}} \delta^{(D-1)}(x-y) \biggr( \delta_{(k}^i\delta_{l)}^j - \frac{1}{D-2} \delta_{(k}^m\delta_{l)}^n h_{mn}(y) h^{ij}(y) \biggr) \\
        =& \frac{16\pi G}{\sqrt{h(y)}} \delta^{(D-1)}(x-y) \biggr( \delta_{(k}^i\delta_{l)}^j - \frac{1}{D-2} h_{kl}(y) h^{ij}(y) \biggr). \\
    \end{split}
\end{equation}

We then need only combine this with a computation of derivatives of $A[\gamma, g]$ with respect to the induced metric.    Proceeding in steps, we introduce the induced metric $q_{AB}$ on $\gamma$ and  $D-2$ coordinates $w^A$ on $\gamma$ to write
\begin{equation}
    A_{HRT}=\int_{\gamma} d^{D-2}w \sqrt{q(w)}.
\end{equation}
Taking functional derivatives yields
\begin{equation}\label{eq:delAdelh}
    \frac{\delta A_{HRT}}{\delta h_{kl}(x)} = \frac{1}{2} \int_{\gamma} d^{D-2}w \sqrt{q(w)} q^{AB}(w) \frac{\delta q_{AB}(w)}{\delta h_{kl}(x)}.
\end{equation}

Now, since $q_{AB}(w) = \frac{\partial x^i}{\partial w^A} \frac{\partial x^j}{\partial w^B} h_{ij}(x)$ (with derivatives computed along $\gamma$), we can rewrite $q_{AB}$ as
\begin{equation}
    \begin{split}
        q_{AB}({w}) =& \int_{\gamma} d^{D-2}\tilde w \frac{\partial x^i}{\partial \tilde w^A} \frac{\partial x^j}{\partial \tilde w^B} h_{ij}(x(\tilde w)) \delta_{\gamma}^{(D-2)}(w,\tilde{w}) \\
        =& \int_{\Sigma} d^{D-1}x \frac{\partial x^i}{\partial \tilde w^A} \frac{\partial x^j}{\partial \tilde w^B} h_{ij}(x) \ \delta_{\gamma}^{(D-2)}(w,\tilde{w}(x)) \  \delta_{\Sigma}(\gamma,x)
    \end{split}
\end{equation}
where $\delta_{\gamma}^{(D-2)}(w,\tilde{w}(x))$ is a $\delta$-function on the HRT surface that satisfies $\int_\gamma d^{D-2}w \ f(w) \delta_{\gamma}^{(D-2)}(w,\tilde w) = f(\tilde w)$,  and where $\delta_{\Sigma}(\gamma,x)$ is a $\delta$-function on the Cauchy slice that localizes $x$ to the HRT surface according to $\int_\Sigma d^{D-1}x f(x) \delta_{\Sigma}(\gamma,x) = \int_\gamma d^{D-2}w f(x(w))$.  We have also arbitrarily extended $\tilde w$ and $\frac{\partial x^i}{\partial \tilde w^A}$ to smooth functions of the $x^i$ defined on all of $\Sigma$, though due to the delta-functions the result does not depend on the particular extension chosen.  We thus find
\begin{equation}\label{eq:delqdelh}
    \frac{\delta q_{AB}(w)}{\delta h_{kl}(x)} = \frac{\partial x^k}{\partial \tilde w^A} \frac{\partial x^l}{\partial \tilde w^B} \ \delta_{\gamma}^{(D-2)}(w,\tilde{w}(x)) \ \delta_{\Sigma}(\gamma,x).
\end{equation}
Finally, combining equations \eqref{eq:hKPB}, \eqref{eq:delAdelh} and \eqref{eq:delqdelh} yields
\begin{eqnarray}
\label{eq:AKPB}
     \biggr\{\frac{A_{HRT}}{4G},K^{ij}(x)\biggr\} &=& 2\pi  \frac{\sqrt{q(\tilde{w}(x))}}{\sqrt{h(x)}} \delta_{\Sigma}(\gamma,x) \biggr( q^{AB}(\tilde{w}(x)) \frac{\partial x^i}{\partial \tilde w^A} \frac{\partial x^j}{\partial \tilde w^B} - h^{ij}(x) \biggr) \\
      &=& - 2\pi  \hat\delta_{\Sigma}(\gamma,x) \perp^i \perp^j.
\end{eqnarray}
where $\perp^i$ is the unit normal to $\gamma$ in $\Sigma$ and $\hat\delta_{\Sigma}(\gamma,x) = \frac{\sqrt{q(\tilde{w}(x))}}{\sqrt{h(x)}} \delta_{\Sigma}(\gamma,x)$ is a one-dimensional Dirac delta-function of the proper distance between $x$ and $\gamma$ measured along geodesics in $\Sigma$ orthogonal to $\gamma$.

Equation \eqref{eq:AKPB} is our main result.  Since the Poisson Bracket with $h_{ij}$ vanishes, and since the right-hand-side of \eqref{eq:AKPB} is the same for all solutions when expressed in terms of proper distance, it is easy to integrate the above to yield the effect of a finite flow by a parameter $\lambda$.   We see that the Hamiltonian flow generated by $A_{HRT}$ changes the initial data on any Cauchy surface $\Sigma$ that contains $\gamma$ by adding to the normal-normal component $K^{\perp \perp}$ of the extrinsic curvature a delta-function given by the right-hand-side of \eqref{eq:AKPB} multiplied by $\lambda$, but that this flow leaves unchanged both the induced metric $h_{ij}$ and all other components of $K^{ij}$.

The effect on the rest of the solution is then determined by the equations of motion.  Note that since there is no change in the initial data on $\Sigma$ away from the HRT surface $\gamma_R$, causality then implies that there can be no change in the part of the solution within the entanglement wedge on either side of $\gamma_R$.  Instead, the solution can change only within the past and future light cones of $\gamma_R$.  In these regions, the change in the solution is also influenced by boundary conditions.  We thus now discuss the required boundary conditions in detail.

\subsection{Boundary Conditions for HRT-area flow}
\label{sec:BCs}

The result \eqref{eq:AKPB} fully defines the flow generated by $A_{HRT}$.  However, as is often the case, the precise connection to boundary conditions can be subtle.  We thus take a moment to explore such issues here.

To this end, recall that \eqref{eq:AKPB} describes a flow within some particular notion of the gravitational phase space.  We have described this phase space in terms of a Cauchy surface $\Sigma$.  The bulk geometry and extrinsic curvature of $\Sigma$ are dynamical and so can change under HRT-area flow.  But since $\Sigma$ represents a definite instant of time, in a context with an asymptotically AdS boundary $\partial {\cal M}$ on which the boundary metric has been fixed, the intersection $\partial \Sigma$ of $\Sigma$ with $\partial {\cal M}$ will remain fixed.  This is in precise analogy with the familiar statement that the flow generated by a Hamiltonian on the phase space at $t=0$ does not actually change the value of $t$ but, instead, changes the initial data in the manner dictated by time-translations.  As a result, the boundary conditions require that neither the metric induced on $\partial \Sigma$ by the boundary metric nor the corresponding extrinsic curvature can change under the flow generated by $A_{HRT}$.  And this must be true despite the transformation \eqref{eq:AKPB} of the initial data in the bulk.

The above may at first seem like a paradoxical state of affairs.  However, any relation between the extrinsic curvature of the surface $\Sigma$ in the bulk ${\cal M}$ and the extrinsic curvature of $\partial \Sigma$ in the boundary $\partial {\cal M}$ will certainly depend on how $\partial {\cal M}$ is attached to ${\cal M}$. This allows extra degrees of freedom. In short, we believe that the situation is much like the famous issue discussed in \cite{Marolf:2010tg,Casini:2011kv,Lewkowycz:2013nqa} wherein one may have conical singularities in the bulk that end on smooth boundary metrics. We thus believe that there is an appropriate sense in which HRT-area flow is a well-defined transformation.  Indeed, we will show this explicitly below for spacetimes asymptotic to AdS$_3$, though we leave full discussion of the higher dimensional case for later work.  In particular, the forthcoming work \cite{DMRGFLOW} will show that our issue is precisely equivalent to whether one can have Lorentz-signature bulk conical singularities in the presence of general smooth boundary metrics.

We also pause to warn the reader that, while we believe that HRT-area flow can be defined, there is a sense in which it will be rather singular in the UV.  In particular, we will see in section \ref{sec:AdS3} that in AdS$_3$ it leads to a boundary stress tensor that involves the square of a Dirac delta-function. The transformed solutions will thus have infinite energy. If we are inspired by \cite{Jafferis:2015del} to think of $A_{HRT}/4G$ as the leading semiclassical term in the modular Hamiltonian of the dual CFT state, this UV-divergence is a concrete manifestation of the singular behavior predicted in \cite{Ceyhan:2018zfg} using results in algebraic quantum field theory.  (Though see \cite{DMRJLMS} for further comments.)  As noted in \cite{Ceyhan:2018zfg} (and as further developed in \cite{Bousso:2019dxk,Bousso:2020yxi}), the UV behavior can be improved by simultaneously acting with a second transformation associated with the (right) vacuum modular Hamiltonian.  Following \cite{Bousso:2020yxi}, we refer to the combined smoother transformation as the `kink transform,' whose details we describe below.  See also the closely related discussions in \cite{Jafferis:2014lza} and \cite{Faulkner:2018faa}.

\subsection{Relation to the kink transformation}
\label{sec:kinkrel}

As a brief but useful aside, we now discuss the relation of the flow generated by $A_{HRT}$ to the kink transformation introduced in \cite{Bousso:2020yxi}.  Indeed, the kink transformation was initially defined in \cite{Bousso:2020yxi} by using precisely the action \eqref{eq:AKPB} on Cauchy data, scaled by a factor that controls the amount of the transformation to be applied.\footnote{\label{foot:norm} We will discuss such normalizations in appendix \ref{app:norm}.
Performing a finite transformation by an amount $\lambda$ simply adds $\lambda$ times the left-hand-side of \eqref{eq:AKPB} to the extrinsic curvature.  However, the astute reader will notice that the form of the normalization factor given in \cite{Bousso:2020yxi} is somewhat different. This difference in presentation will be discussed at the end of appendix  \ref{app:norm}.}  The transformation on solutions then followed by solving the equations of motion.  However, for asymptotically AdS spacetimes the solution is unique only after boundary conditions have been fully specified, and the boundary conditions chosen to define the kink transformation in \cite{Bousso:2020yxi}  turn out to differ from the HRT-area flow boundary conditions described in section \ref{sec:BCs}.  While the flow generated by $A_{HRT}$ preserves {\it any} boundary metric and leaves $\partial \Sigma$ invariant in $\partial {\cal M}$, the kink transform of \cite{Bousso:2020yxi} was fully defined only when the metric on $\partial {\cal M}$ has a Killing field $\xi_\partial$ that vanishes on the anchor set $\partial R$ of the HRT surface, and where $\xi_\partial$ acts locally as a boost about $\partial R$.  In this setting the kink transformation was declared to leave the boundary metric invariant, and also to leave the surface $\partial \Sigma$ invariant in the region spacelike separated from $R$.  However, in contrast to the HRT-area flow described above, the kink transformation moves the part of $\partial \Sigma$ in the domain of dependence of $R$.  In particular, it shifts $\partial \Sigma$ toward the past along the orbits of $\xi_\partial$ by a Killing parameter $2\pi \lambda/\kappa$, where $\kappa$ is the surface gravity of $\xi_\partial$ at $\partial R$.  In all cases below we take $D(R)$ to be the right wedge and describe the left wedge as $D(\bar R)$ for some complimentary achronal surface $\bar R$ to $R$.  See appendix \ref{app:norm} for verification of the above sign and normalization factors.

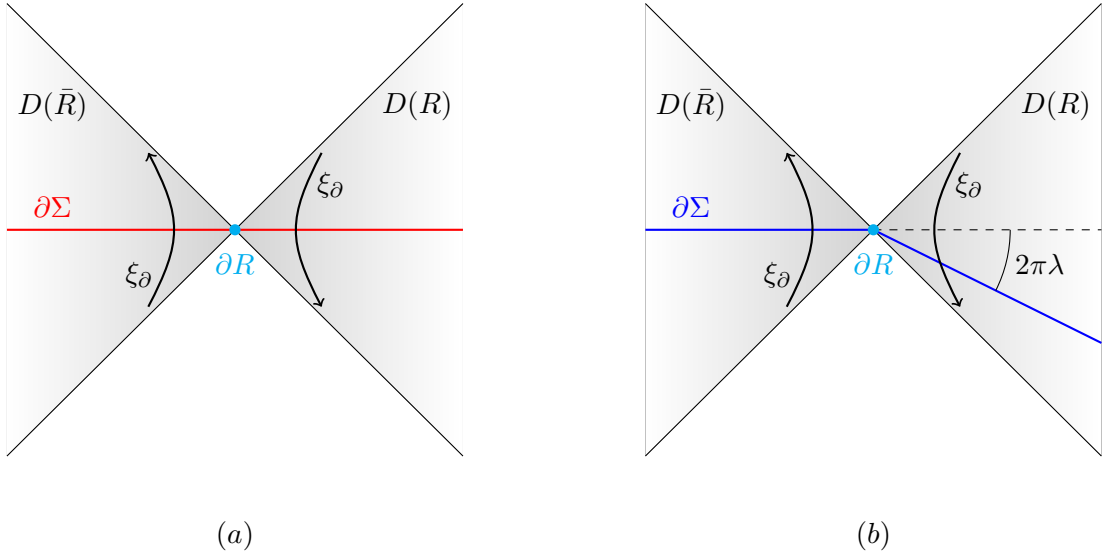
\begin{figure}
    \centering
    \begin{tikzpicture}[scale=3]
    \fill[left color=white,right color=lightgray!70] (-1,-1) -- (0,0) -- (-1,1);
    \fill[right color=white,left color=lightgray!70] (1,1) -- (0,0) -- (1,-1);
    \draw (-1,1) -- (1,-1);
    \draw (-1,-1) -- (1,1);
    \draw[red,thick] (-1,0) -- (1,0);
    \draw[red] (-.8,.1)  node {$\partial \Sigma$};
    \draw (0,-1.35) node {$(a)$};
    \fill[cyan] (0,0) circle (.7pt);
    \draw[cyan] (0,-.15) node {$\partial R$};
    \draw (-.8,.55) node {$D(\bar{R})$};
    \draw (.8,.55) node {$D(R)$};
    \draw (.42,.2)  node {$\xi_{\partial}$};
    \draw (-.42,-.2)  node {$\xi_{\partial}$};
    \draw[->, thick] (.38,.34) .. controls (.23,.05) and (.23,-.05) .. (.38,-.34);
    \draw[<-, thick] (-.38,.34) .. controls (-.23,.05) and (-.23,-.05) .. (-.38,-.34);

\begin{scope}[xshift=2.8cm]
    \fill[left color=white,right color=lightgray!70] (-1,-1) -- (0,0) -- (-1,1);
    \fill[right color=white,left color=lightgray!70] (1,1) -- (0,0) -- (1,-1);
    \draw (-1,1) -- (1,-1);
    \draw (-1,-1) -- (1,1);
    \draw[blue,thick] (-1,0) -- (0,0) -- (1,-.5) ;
    \draw[dashed] (0,0) -- (1,0);
    \draw (.42,.2)  node {$\xi_{\partial}$};
    \draw[blue] (-.8,.1)  node {$\partial \Sigma$};
    \draw (-.42,-.2)  node {$\xi_{\partial}$};
     \draw (.6,0) arc[start angle=0, delta angle=-26.57, radius=0.6cm]; 
    \draw (.73,-.15) node {$2\pi \lambda$};
    \draw (0,-1.35) node {$(b)$};
    \draw[->, thick] (.38,.34) .. controls (.23,.05) and (.23,-.05) .. (.38,-.34);
    \draw[<-, thick] (-.38,.34) .. controls (-.23,.05) and (-.23,-.05) .. (-.38,-.34);
    \fill[cyan] (0,0) circle (.7pt);
    \draw[cyan] (0,-.15) node {$\partial R$};
    \draw (-.8,.55) node {$D(\bar{R})$};
    \draw (.8,.55) node {$D(R)$};
\end{scope}
\end{tikzpicture}
    \caption{The conformal boundary of our spacetime, showing the domains of dependence $D(R)$ of $R$ and $D(\bar R)$ of $\bar R$.  The boundary metric has a Killing field $\xi_\partial$ that acts as a boost near $\partial R$.  {\bf (a)}  The boundary $\partial \Sigma$ of a smooth bulk Cauchy surface $\Sigma$ in the original spacetime.  The surface $\partial \Sigma$ and all boundary observables on that surface are preserved by the flow generated by $A_{HRT}{R}$. {\bf (b)} The kink transformation with parameter $\lambda$ moves the part of $\partial \Sigma$ in $D(R)$ by sliding it toward the past along orbits of $\xi_\partial$ through a Killing parameter $2\pi\lambda/\kappa$.  Near $\partial R$ this acts as a past-directed boost with rapidity $2\pi \lambda$. }
    \label{fig:bndy_actions}
\end{figure}

In the presence of the boundary Killing field $\xi_\partial$, the kink transformation differs from the flow generated by $A_{HRT}/4G$ only by whether or not $\partial \Sigma$ is distorted relative to the fixed boundary metric. We may thus refer to the flow generated by $A_{HRT}$ as a {\it boundary-condition preserving} kink transform.  Again, because this flow preserves the boundary conditions precisely, it can be defined for any boundary metric.  In particular, it does not require the existence of the boundary Killing field $\xi_\partial$ that was needed to define the original kink transform.

Since the above distortion involves a boost operation in the right ($R$) wedge but trivial action in the left ($\bar R$) wedge, it was called a (boundary) one-sided boost in \cite{Bousso:2020yxi}.  Note, however, that the action on observables in the future and past wedeges is again determined by solving the equations of motion.      If we let ${\cal K}[\gamma]$ denote the generator of the kink transformation by $\lambda$, then we can define the difference $H_R : = \frac{A_{HRT}}{4G} -{\cal K}[\gamma]$ to be ($2\pi$ times) the generator of the boundary one-sided boost (taken to generate flow toward the future in the right wedge).  In the context of AdS/CFT, $H_R$ can be interpreted \cite{Bousso:2020yxi} as  the right modular Hamiltonian of the Hartle-Hawking state\footnote{There will be cases where $H_R$ is unbounded below as a CFT operator.  In such cases the Hartle-Hawking state is not well-defined, but the flow still exists.  Such cases are the analogue of what occurs for Kerr black holes in asymptotically flat spacetimes.} for the CFT associated with the boundary Killing field $\xi_\partial.$    Furthermore, the term $\frac{A_{HRT}}{4G}$ was argued in \cite{Bousso:2020yxi} to correspond at leading order to the modular Hamiltonian of the boundary dual of the bulk spacetime.  As a result, the kink transform was conjectured to be dual to a so-called Connes cocycle flow in the CFT (generated by the difference between the right modular Hamiltonian of the bulk state and the right modular Hamiltonian of the Hartle-Hawking state for $\xi_\partial$).  Some refinements of this correspondence will be discussed in \cite{DMRJLMS}.

It is useful to note that, even in the absence of a bulk Killing field, the action of $A_{HRT}$ or ${\cal K}[\gamma]$ in the bulk can again be described in terms of a one-sided boost.  This relationship was described in detail in \cite{Bousso:2020yxi}, having been foreshadowed in \cite{Jafferis:2014lza,Ceyhan:2018zfg,Faulkner:2018faa,Bousso:2019dxk}.   The essential point is to recall from \cite{Engelhardt:2018kcs} that the original solution can be reconstructed from four pieces of data: boundary conditions as defined by the boundary metric, the restriction of the solution to the left wedge, the restriction of the solution to the right wedge, and the way that affine parameters along the future and past null boundaries of each wedge are identified with those along the past and future null boundaries of the other wedge.  The idea is that if we are given the last three, the remainder of the solution is uniquely determined by solving the equations of motion subject to the given boundary conditions (the first ingredient above).

The fourth piece of data above can be said to define the relative boost with which the two wedges are attached.  The desired operation is then defined by changing these identifications in precisely the same way that they would be changed if there were an appropriate bulk Killing field, and if we were to transform the right wedge by flowing toward the past through a Killing parameter $2\pi \lambda/\kappa$ along the orbits of this Killing field.  As verified in appendix \ref{app:norm}, on a Cauchy surface through $\gamma$ this generates precisely the desired transformation on initial data \eqref{eq:AKPB}.  

Again, the transformed initital data can be extended to a full solution by choosing boundary conditions and solving the equations of motion.  And again, the result gives the flow generated by either $A_{HRT}/4G$ (if one preserves the way that each wedge attaches to the asymptotically AdS boundary), or by ${\cal K}[\gamma]$ (if there is a boundary Killing field $\xi_\partial$ and one flows the right wedge appropriately under $\xi_\partial$).  In all cases  the solution in the past and future of $\gamma$ is determined by solving the equations of motion with an appropriate choice of boundary conditions.\footnote{In particular, since the boundary metric is not dynamical, the boundary metric to the future and past of $\partial R$ cannot be determined by solving equations of motion.  It must simply be specified by hand.}

\section{Explicit results in vacuum Poincar\'e AdS$_3$}
\label{sec:AdS3}

We will now use the above relations to give a simple geometric description of the flow generated by some $A_{HRT}[R]$ in pure 2+1 Einstein-Hilbert gravity (with negative cosmological constant but without matter) for spacetimes asymptotic to Poincar\'e AdS$_3$ that do not contain black holes.  After deriving this description in section \ref{sec:BCTR}, explicit results for the action of the transformation on the boundary stress tensor and on other HRT-areas are given in sections \ref{sec:AdS3Tijact} and \ref{sec:HRTAact}.

\subsection{Representation as a boundary conformal transformation}
\label{sec:BCTR}

Bulk spacetimes of the specified form are always diffeomorphic to Poincar\'e AdS$_3$. Let us thus focus on obtaining explicit results when the spacetime is exactly Poincar\'e AdS$_3$ with metric
\begin{equation}
\label{eq:PAdS3}
ds^2 = \frac{1}{z^2}\left( -dt^2 + dx^2 + dz^2\right) = \frac{1}{z^2}\left( -dudv + dz^2\right).
\end{equation}
Here we have set the AdS scale $\ell_{AdS}$ to one and introduced $u=t-x$ and $v=t+x$.
Results for any other spacetime in the above class can then be obtained by applying an appropriate boundary conformal transformation.  At least for infinitesimal such transformations, this generalization will be described in section \ref{sec:conf}.

Now, any two HRT surfaces in Poincar\'e AdS$_3$ are related by an AdS$_3$ isometry. Thus we may further simplify the discussion by taking the boundary region $R$ to be  the half-line $x\in[0, \infty)$ at $t=0$ on the boundary at $z=0$.  We will refer to this half-line as $R_0$.  The HRT surface $\gamma_{R_0}$ is then the bulk geodesic given by $x=t=0$ for all $z$.

The geodesic $\gamma_{R_0}$ is invariant under the manifest boost isometry $\xi = x \partial_t + t \partial_x$ in the $x,t$ plane, and it is clear that $\xi$ induces a related Killing field $\xi_\partial$ on the boundary at $z=0$.  This feature makes it easy to apply the kink transformation ${\cal K}[\gamma_{R_0}]$, as boosting the right wedge leaves invariant {\it all} data in that wedge.  The kink transformation also leaves the boundary metric unchanged, though we remind the reader that it nevertheless `moves each Cauchy surface with respect to that metric' as shown previously in figure \ref{fig:bndy_actions}.  As a result, solving the equations of motion must  precisely reproduce the original spacetime \eqref{eq:PAdS3}.  We conclude that the action of ${\cal K}[\gamma_{R_0}]$ leaves Poincar\'e AdS$_3$ invariant.\footnote{This is consistent with the conjecture of \cite{Bousso:2020yxi} that the kink transform is dual to the Connes cocycle flow generated by the difference between the one-sided modular Hamiltonian of the dual CFT state and the one-sided modular Hamiltonian of the CFT vacuum. Since Poincar\'e AdS$_3$ is dual to the CFT vacuum, the above difference clearly vanishes for this state and hence has trivial action.  As usual, we refer the reader to \cite{DMRJLMS} for further comments}

A similar conclusion clearly holds for Poincar\'e AdS$_d$ for any $d$.  But what is special about $d=3$ is that we can also find a simple form for the transformation generated by the $H_R$ of section \ref{sec:kinkrel}.  Combining this with the above will then give a closed-form expression for the boundary-condition-preserving kink transformation defined by our HRT-area flow  on Poincar\'e AdS$_3$.

To establish the desired result, recall first that $H_R$ generates a transformation that leaves invariant the boundary metric. And since all solutions to pure 2+1 Einstein-Hilbert gravity with such boundary conditions are diffeomorphic to Poincar\'e AdS$_3$, $H_R$ can act only by a boundary-metric-preserving diffeomorphism. In an asymptotically AdS spacetime, this must be a boundary conformal transformation.  Our task is thus simply to identify the unique conformal transformation that acts on the right wedge of the boundary as a boost of magnitude $2 \pi \lambda$ and of the appropriate sign.\footnote{While this conformal transformation is not smooth on the boundary spacetime, it nevertheless corresponds to a diffeomorphism that is smooth at every point in the bulk.}

For each $\lambda$ we will describe this conformal transformation as a map $(u,v) \rightarrow (U(u),V(v))$ on the boundary spacetime and an associated Weyl rescaling. After acting with the transformation,  our boundary conditions require the boundary metric to be
\begin{equation}
\label{eq:bndymetric}
ds^2_\partial =  - dU dV,
\end{equation}
so that the Weyl rescaling is determined by comparing \eqref{eq:bndymetric} with $-dudv$.

In the left wedge we know that $H_R$ must act as the identity.  And in order to undo the action of ${\cal K}[\gamma]$ on a Cauchy slice (shown in figure \ref{fig:bndy_actions}), our conformal transformation should boost the right wedge toward the future with rapidity $2\pi \lambda$.  Since it must preserve continuity of each Cauchy slice, this uniquely singles out the transformation at each finite $\lambda$ to be

\begin{equation}
\label{eq:bctm}
U = u e^{-2\pi \lambda \Theta(-u)},  \ \ \ V = v e^{2 \pi \lambda \Theta(v)}.
\end{equation}
Since $u\delta(u) = 0 = v \delta(v)$, \eqref{eq:bctm} yields
\begin{eqnarray}
 && -  dUdV = - e^{2 \sigma_- (U)}e^{2 \sigma_+ (V)} du dv \\
 \label{eq:uvbm}
{\rm with} \ &&  e^{2 \sigma_- (U)} = e^{-2 \pi\lambda \Theta(-U)} , \ \ \ e^{2 \sigma_+ (V)} = e^{2 \pi \lambda \Theta(V)}.
\end{eqnarray}

Thus we have $-dUdV = - dudv$ in both the left and right wedges.  But this is not the case in either of the future or past wedges, so  the $\sigma_\pm$ define a non-trivial Weyl rescaling relating \eqref{eq:uvbm} to \eqref{eq:bndymetric}.

On any solution, $H_R$ will be the generator of the boundary conformal transformation \eqref{eq:bctm}.  But since the kink transform acts trivially on Poincar\'e AdS$_3$,
we can also take \eqref{eq:bctm} to give the full finite-$\lambda$ action flow of this solution under $A_{HRT}/4G$.
This in particular allows us to explicitly compute the action of this flow on both the boundary stress tensor and other HRT areas.  We record these results below in sections \ref{sec:AdS3Tijact} and \ref{sec:HRTAact} for later use in comparison with section \ref{sec:conf}.

\subsection{HRT-area flow of $T_{ij}$}
\label{sec:AdS3Tijact}

The action of a general finite conformal transformation on the stress energy tensor of a 1+1 dimensional conformal field theory is well known (see e.g. \cite{DiFrancesco:1997nk}) to give

\begin{equation}\label{eq:T}
    T_{ab}dx^a dx^b = T_{ab}^{\text{original}}dx^a dx^b + \frac{c}{12\pi}\biggr[\partial_U^2\sigma_- + (\partial_U \sigma_-)^2\biggr]dU^2 + \frac{c}{12\pi}\biggr[\partial_V^2\sigma_+ + (\partial_V \sigma_+)^2\biggr]dV^2,
\end{equation}
where $c$ is the central charge. For the boundary stress tensor of AdS$_3$, we have $c= 3/2G$. Since we are computing the effect on the planar vacuum, we have  $T_{ab}^{\text{original}}=0$. The remaining terms in \eqref{eq:T} then give
\begin{eqnarray}
\label{eq:AtransTijU}
T_{UU} &=& \frac{1}{8 G} \left( \lambda \delta'(U)  + \pi \lambda^2 [\delta(U)]^2 \right), \ {\rm and} \\
\label{eq:AtransTijV}
T_{VV} &=& \frac{1}{8 G} \left(  \lambda \delta'(V)   +  \pi \lambda^2 [\delta(V)]^2 \right).
\end{eqnarray}

The final terms in \eqref{eq:AtransTijU} and \eqref{eq:AtransTijV} are sensible only in the presence of a UV regulator.  This is consistent with comments in \cite{Ceyhan:2018zfg} on the singular nature of  one-sided modular flow.  Interestingly, however, there is no problem at linear order in $\lambda$.
This makes clear that the infinitesimal action of HRT-area flow is well-defined on solutions that are sufficiently smooth, but that flowing a finite distance under this transformation creates UV divergences when the HRT surface $\gamma$ is anchored to an asymptotically AdS boundary.  In all cases we nevertheless emphasize that the action of the flow on the bulk solution is nevertheless given in closed form.

\subsection{The action of the flow on other HRT-areas}
\label{sec:HRTAact}

Despite the divergence it creates in the boundary stress tensor components \eqref{eq:AtransTijU} and \eqref{eq:AtransTijV}, the finite flow generated by $A_{HRT}[R_0]/4G$ yields a well-defined action on other HRT areas.  To write explicit formulae, recall  that our $A_{HRT}$ denotes the {\it renormalized} HRT-area, which in AdS$_3$ with $\ell_{AdS}=1$ may be written
\begin{equation}
\label{eq:renormA}
        A_{HRT} = L_{geodesic} + \sum_{anchors} L_{ct},
\end{equation}
where $L_{ct}$ is an appropriate $c$-number counterterm.
In particular, in vacuum Poincar\'e AdS$_{3}$ we may introduce a  regulated boundary at $z=\epsilon$ and write the renormalized area as \cite{Morrison_2013}
\begin{equation}
\label{eq:AHRTvacexp}
    \begin{split}
       A_{HRT}^{vac} =& \lim_{\epsilon \to 0} [-2\ln \epsilon + \ln[(x_1-x_2)^2-(t_1-t_2)^2] + 2\ln(2\epsilon)] \\
       =& \ln [ (u_1-u_2)(v_2-v_1)]  + 2\ln 2,
    \end{split}
\end{equation}
where we have identified $L_{ct} = \ln(2\epsilon)$.  Since the endpoints of the HRT surface must be spacelike separated on the boundary, without loss of generality we may take $u_1 > u_2$ and $v_1 < v_2$ (i.e., we number the endpoints  left-to-right as opposed to past-to-future).

Furthermore,
under a Weyl rescaling $ds^2_{\partial, new}  = e^{2\sigma} ds^2_{\partial, old}$ we have
\begin{equation}\label{eq:Anew}
        A_{HRT}^{new} = A_{HRT}^{old} +  \sum_{anchors} \sigma.
\end{equation}

We can now apply the conformal transformation \eqref{eq:uvbm} to the $A_{HRT}$ anchored at $(U_1,V_1)$ and $(U_2,V_2)$, which we write below as $A_{HRT}(U_1,V_1,U_2,V_2)$.  First, however, it is useful to note that any $A_{HRT}$ evaluated in the Poincar\'e vacuum remains invariant under the conformal transformation defined by constant rescalings $U= e^{\alpha_u} u, V = e^{\alpha_v}v$ of the null coordinates.  This is because the explicit expression \eqref{eq:AHRTvacexp} shifts by $-\alpha_u - \alpha_v$ under $(u,v) \rightarrow (U,V)$, but this is then cancelled by the conformal anomaly term in \eqref{eq:Anew} since $\sigma = \frac{1}{2}(\alpha_u + \alpha_v)$ at each anchor point.  This result is of course clear for the case $\alpha_u = - \alpha_v$ (which describes a boost), but it also holds for e.g. $\alpha_u = \alpha_v$ (which describes a dilation).

Since \eqref{eq:uvbm} is piecewise constant, the above observation makes it easy to apply the conformal transformation
\eqref{eq:bctm}  to $A_{HRT}(U_1,V_1,U_2,V_2)$. Non-trivial effects can occur only when $U_1$ and $U_2$ have opposite signs or when $V_1,V_2$ have opposite signs so that the end points correspond to distinct values of $\sigma_+$ and/or $\sigma_-$.  When opposite signs do occur, we can evaluate the effect of \eqref{eq:bctm} by computing \eqref{eq:AHRTvacexp} at the new endpoints and adding the anomalous term from \eqref{eq:Anew}.   Again using spacelike separation of the anchor points to take $U_1 > U_2$ and $V_1 < V_2$, we may write the transformed result in the form
\begin{equation}
\label{eq:AHRTlamvacexp}
A_{HRT,\lambda}(U_1,V_1,U_2,V_2) = A_U(U_1,U_2) + A_V(V_1,V_2) + 2 \ln 2,
\end{equation}
with
\begin{equation}
\label{eq:AHRTlamvacexpU}
A_U =            \begin{cases} \ln(U_1-U_2),  & U_1,U_2<0 \ {\rm or} \ U_1,U_2 > 0\\
\ln(e^{-2\pi \lambda} U_1-U_2),  & U_2 < 0 < U_1
\end{cases}
\end{equation}
and
\begin{equation}
\label{eq:AHRTlamvacexpV}
A_V =            \begin{cases} \ln(V_2-V_1),  & V_1,V_2<0 \ {\rm or} \ V_1,V_2 > 0\\
\ln(V_2 - e^{2 \pi \lambda} V_1),  & V_1 < 0 < V_2.
\end{cases}
\end{equation}

As a result, we find
\begin{eqnarray}\label{eq:AAgeo}
\biggr\{ \frac{1}{4G}A_{HRT}[R_0],  A_{HRT}(U_1,V_1,U_2,V_2) \biggr\} &=& \frac{d}{d\lambda}  A_{HRT, \lambda}(U_1,V_1,U_2,V_2)|_{\lambda =0} \\ &=& -2\pi \left( \frac{U_1 \Theta(-U_1U_2)}{U_1 - U_2} + \frac{V_1 \Theta(-V_1V_2)}{V_2 - V_1}\right) \label{eq:AAgeo2}.
\end{eqnarray}
%%%%%%%%%%%%%%%%%%%%%%%%%%%%%%%%%%%%%%%%%%%%%

%%%%%%%%%%%%%%%%%%%%%%%%%%%%%%%%%%%%%%%%%%%%%%%%%%%%%%%%%%%%%%%%%%%%%%%%%%%%%%%%%%%
\section{Commutators from stress tensors}\label{sec:conf}

The previous section transformed the general arguments of section \ref{sec:kink} into explicit results in AdS$_3$ for the particular choice of boundary region $R_0$ given by the half-line at $t=0$ and when the area-flow acts on the Poincar\'e AdS$_3$ vacuum.  We now present an independent calculation both to check the above results and as a means of generalizing them to allow arbitrary boundary regions $R$ and more general solutions of pure 2+1 Einstein-Hilbert gravity with negative cosmological constant and planar boundary.  In particular, the generalization will allow planar black holes.  The key ingredients in this computation are (again) that all such solutions are related by boundary conformal transformations, and that any two spacetimes with the same boundary stress tensor are considered to be completely equivalent.  Thus we may in principle express any observable in terms of the
boundary stress tensor and use the well-known 2-dimensional stress tensor algebra to compute any commutators.  This alternate technique may also be of interest in its own right as a means of studying commutators of general quantities for which an elegant geometric description of the flow is not known.

To follow this approach, one might like to proceed by finding an explicit expression in terms of the boundary stress tensor for the conformal transformation $(u,v) \rightarrow (U(u),V(v))$ that constructs an arbitrary solution in our class from the Poincar\'e AdS$_3$ vacuum, and in particular for the associated conformal factor  $\sigma[T_{ij}]$.   One could then use the relevant conformal anomaly to write any observable in terms of $\sigma$ and simply substitute $\sigma[T_{ij}]$ to write the observable as a functional of the stress tensor.  However, it is not clear that a useful such closed-form solution $\sigma[T_{ij}]$ will exist.  Luckily this work focusses on {\it semi-classical} commutators, for which such an explicit relation will not be needed.  As in section \ref{sec:kink}, the key point is that the semiclassical commutator (aka the Poisson or Peierls bracket) is a derivation, meaning that for any observables $B$ and $C$ we may write the bracket in terms of the stress-tensor algebra using
\begin{equation}\label{eq:leibniz}
\{B,C\} = \int d^2x_1 d^2x_2 d^2x'_1 d^2x'_2 \frac{\delta B}{\delta \sigma(x_1)} \frac{\delta \sigma(x_1)}{\delta T_{ij}(x_2)} \{T_{ij}(x_2),T_{i'j'}(x'_2) \} \frac{\delta \sigma(x_1')}{\delta T_{i'j'}(x'_2)} \frac{\delta C}{\delta \sigma(x_1')}.
\end{equation}
It is thus sufficient to know the functional derivatives of $B,C$ with respect to $\sigma$ and the functional derivatives of $\sigma$ with respect to the stress tensor, which in practice turns out to be a manageable task.

After computing such functional derivatives in section \ref{sec:sigT1}, we warm up with some relatively simple commutators involving $\sigma$ and $T_{ij}$ in sections \ref{sec:sigT} and \ref{sec:sigsig} before finally studying commutators of HRT areas in section \ref{sec:area}.

\subsection{Functional derivatives of $\sigma$ with respect to $T_{ij}(x)$}
\label{sec:sigT1}

Recall that our basic strategy will be to think of the boundary stress tensor $T_{ij}(U,V)$ as the fundamental observable in terms of which we will write all others.  In particular, we will define an observable $\sigma$ as the conformal factor in \eqref{eq:T} that generates $T_{ij}(U,V)$ from the vacuum (in which $T_{ij}=0$), and which satisfies certain boundary conditions.  In practice, we will then write general observables in terms of $\sigma$, which then implicitly expresses them in terms of $T_{ij}$.

To this end, let us thus consider the boundary metric
\begin{equation}
    ds^2_\partial = - dU dV =-e^{2\sigma(u,v)}dudv,
\end{equation}with $\sigma(u,v) = \sigma_-(u) + \sigma_+(v)$ and
\begin{equation}
\label{eq:uUvV}
    \begin{split}
        dU &= e^{2\sigma_-(u)}du \\
        dV &= e^{2\sigma_+(v)}dv,
    \end{split}
\end{equation}
and where $\sigma_\pm$ are chosen so that $T_{ij}(U,V)$ can be written in the form \eqref{eq:T} with 
$T^{original}_{uu}(u) = 0 = T^{original}_{vv}(v)$.  Note that the coordinates $u,v$ defined by \eqref{eq:uUvV} are dynamical objects that will also be functions of the stress tensor $T_{ij(U,V)}$.  In particular, we should think of the functions $u(U)$ and $v(V)$ as being defined by integrating the above equations subject to some boundary condition.  These functions have range $u \in (-\infty,u_{max})$ and $v \in (-\infty,v_{max})$, where $u_{max}=v_{max}=\infty$ when solutions asymptote to Poincar\'e AdS$_3$, and $u_{max}=v_{max}=0$ when solutions instead asymptote to an $M>0$ planar black hole.  

When solutions asymptote to Poincar\'e AdS$_3$, it will be convenient to choose the boundary conditions to be simply
\begin{equation}
\label{eq:UVBC}
u(U=0) =0, \ \ \ v(V=0) =0,
\end{equation}
allowing us to define
\begin{eqnarray}
    u(U) =& \int_{0}^U dU' e^{-2\sigma_-(U')} \\
    v(V) =& \int_{0}^V dV' e^{-2\sigma_+(V')}.
\end{eqnarray}
We note in passing that $u=\pm \infty$ does not generally correspond to $U=\pm \infty$.  In particular, at this stage the $SL(2,R) \times SL(2,R)$ isometries of the Poincar\'e vacuum allow us to introduce poles in the function $u(U)$.
When solutions asymptote to an $M>0$ planar black hole, we will instead choose our boundary conditions to be
\begin{equation}
\label{eq:UVBCBH}
u(U=\infty) =0, \ \ \ v(V=\infty) =0,
\end{equation}
allowing us to define
\begin{eqnarray}
    u(U) =& \int_{0}^U dU' e^{-2\sigma_-(U')} - \int_0^{\infty}dU' e^{-2\sigma_-(U')} \\
    v(V) =& \int_{0}^V dV' e^{-2\sigma_+(V')} - \int_{0}^{\infty} dV' e^{-2\sigma_+(V')}.
\end{eqnarray}
Again we note in passing that $u=-\infty$ may not correspond to $U=-\infty$.

Since the trace of $T_{ij}$ vanishes, the only non-trivial stress tensor components are $T_{UU}(U)$ and $T_{VV}(V)$.  For the moment, let us focus on $T_{UU}$ since corresponding results for $V$ will satisfy analogous expressions.

The functional derivative $\frac{\delta \sigma_- (u(U))}{\delta T_{UU}(U')}$ can be computed by studying the variation of $T_{UU}$:
\begin{equation}
    \begin{split}
        \delta T_{UU} =& \frac{c}{12\pi}(\partial_U^2\delta\sigma_- + 2\partial_U \sigma_- \partial_U \delta \sigma_-) \\
        =& \frac{c}{12\pi}e^{-2\sigma_-}\partial_{U}(e^{2\sigma_-}\partial_{U}\delta \sigma_-). \\
    \end{split}
\end{equation}
Solving for $\delta \sigma_-(U)$ yields
\begin{equation}\label{eq:delsig}
    \begin{split}
        \delta \sigma_-(U) = \frac{12\pi}{c}\biggr( \int_{U_0}^U e^{-2\sigma_-(U'')} \int_{U_0}^{U''} &e^{2\sigma_-(U')} \delta T_{UU}(U') dU' dU'' \\
        &+ c_1 \int_{U_0}^U e^{-2\sigma_-(U')} dU' + c_2 \biggr),
    \end{split}
\end{equation}
for any finite $U_0$, and constants $c_1$ and $c_2$.  This parametrization is somewhat redundant, as changes in $U_0$ can be absorbed into changes in $c_1$ and $c_2$.

The constants $c_1$ and $c_2$ are arbitrary and cannot influence the physics of our computation.  But it will be convenient to fix them by recalling that,
as previously noted, an overall scaling preserves the Poincar\'e vacuum.  We are thus free to fix $\sigma_-(U)$ to be independent of the stress tensor for any one value $U$.  We will choose this to be true  at $U=U_0$, which imposes $\delta\sigma_-(U_0)=0$.  Additionally, the Poincar\'e vacuum is invariant under special conformal transformations, allowing us to fix  $\partial_U \sigma_-(U)$ to be independent of the stress tensor for any one value $U$; we will again choose $U=U_0$, which imposes $\partial_U \delta \sigma_-(U_0)=0$.  Implementing these boundary conditions sets $c_1=c_2=0$, and establishes that our definition of $\sigma$ depends on $U_0$.  As a result, we now change notation to $\delta\sigma_-(U)=\delta\sigma_{U_0}(U)$, to make explicit the $U_0$ dependence. We similarly take $\sigma_{-}(U_0)=0$.

Rewriting \eqref{eq:delsig} slightly, we have

\begin{equation}
\begin{split}
    \delta \sigma_{U_0}(U) =& \frac{12\pi}{c} \int_{-\infty}^{\infty} dU' e^{2\sigma_{U_0}(U')} \Theta(U-U')\Theta(U'-U_0)  \delta T_{UU}(U') \int_{U'}^{U} dU''  e^{-2\sigma_{U_0}(U'')} \\
    &+ \frac{12\pi}{c} \int_{-\infty}^{\infty} dU' e^{2\sigma_{U_0}(U')} \Theta(U'-U)\Theta(U_0-U')  \delta T_{UU}(U') \int_{U}^{U'} dU''  e^{-2\sigma_{U_0}(U'')},
\end{split}
\end{equation}
which yields
\begin{equation}\label{eq:delsdelT}
    \frac{\delta \sigma_{U_0}(U)}{\delta T_{UU}(U')} = \frac{12\pi}{c}  e^{2\sigma_{U_0}(U')}[u(U)-u(U')][\Theta(U-U')\Theta(U'-U_0)-\Theta(U'-U)\Theta(U_0-U')].
\end{equation}
We now use the above results to compute a series of semiclassical commutators below.
%%%%%%%%%%%%%%%%%%%%%%%%%%%%%%%%%%%%%%
\subsection{A first warm up: $\{\sigma, T_{ij}\}$ }\label{sec:sigT}
We begin by studying the semiclassical commutator
\begin{equation}\label{eq:sigT}
    \{\sigma_{U_0}(U),T_{UU}(\tilde{U})\} = \int_{-\infty}^{\infty} dU' \frac{\delta \sigma_{U_0}(U)}{\delta T_{UU}(U')} \{T_{UU}(U'),T_{UU}(\tilde{U})\}.
\end{equation}
Since our CFT is 2-dimensional, the stress tensor algebra can be determined from the familiar relations
\begin{equation}
\label{eq:nearVir}
    \{L_m,L_n\} = i(n-m)L_{m+n} - \frac{ic}{12}m(m^2-1)\delta_{m+n,0},
\end{equation}
where $L_m$ and $T_{UU}$ are related by
\begin{equation}
    \begin{split}
        L_m &= -\frac{1}{2\pi} \int_{S^1} dU e^{iUm}T_{UU}(U) \\
        T_{UU}(U) &= \sum_{m=-\infty}^{\infty} e^{-iUm}L_m.
    \end{split}
\end{equation}
These relations are well-known to yield
\begin{equation}
\label{eq:sta}
    \{T_{UU}(U),T_{UU}(U')\}= 2T_{UU}(U')\delta'(U-U')-T_{UU}'(U')\delta(U-U') - \frac{c}{24\pi}\delta^{'''}(U-U').
\end{equation}

We can now compute the right-hand side of \eqref{eq:sigT} using \eqref{eq:sta}.
After some manipulation, \eqref{eq:sigT} becomes
\begin{equation}
    \begin{split}
        \{\sigma_{U_0}(U),T_{UU}(\tilde{U})\} = \biggr[ -&2T_{UU}(\tilde{U}) \partial_{U'}\biggr(\frac{\delta \sigma_{U_0}(U)}{\delta T_{UU}(U')} \biggr) - T_{UU}'(\tilde{U})\frac{\delta \sigma_{U_0}(U)}{\delta T_{UU}(U')} \\
        &+ \frac{c}{24\pi}\partial_{U'}^3\biggr(\frac{\delta \sigma_{U_0}(U)}{\delta T_{UU}(U')} \biggr)\biggr]_{U'=\tilde{U}},
    \end{split}
\end{equation}
where we have used \eqref{eq:delsdelT} to integrate by parts and to show that the associated boundary terms vanish at $U=\pm \infty$.  Using Eq.~\eqref{eq:delsdelT} then yields
\begin{equation}\label{eq:sigTFinal}
\begin{split}
    \{\sigma_{U_0}(U),T_{UU}(\tilde{U})\} =& \sigma_{U_0}'(\tilde{U})\delta(U-\tilde{U})- \frac{1}{2} \delta'(U-\tilde{U})\\
    &-\sigma_{U_0}''(U_0)[u(U)-u(U_0)]\delta(\tilde U - U_0)-\sigma_{U_0}'(U_0)\delta(\tilde U - U_0) \\
    &+\sigma_{U_0}'(U_0)[u(U)-u(U_0)]\delta'(\tilde U - U_0) - \frac{1}{2}\delta'(\tilde U - U_0) \\
    &+\frac{1}{2}[u(U)-u(U_0)]\delta''(\tilde U - U_0).
\end{split}
\end{equation}
We will use this result to calculate the commutator $\{\sigma(X), \sigma(X')\}$ in the next section, and we will use it again in Section \ref{sec:AT} to calculate the commutator between an HRT-area and the stress-energy tensor.

%%%%%%%%%%%%%%%%%%%%%%%%%%%%%%%%%%%%
\subsection{The $\sigma$ commutator}\label{sec:sigsig}

Our next step will be to compute $\{\sigma(X), \sigma(X')\}$.
By the Leibniz rule as expressed in Eq.\eqref{eq:leibniz}, we have
\begin{equation}\label{eq:sigsig}
    \{\sigma_{U_0}(U),\sigma_{\tilde U_0}(\tilde{U})\} = \int_{-\infty}^{\infty} d\tilde{U}'  \frac{\delta \sigma_{\tilde U_0}(\tilde{U})}{\delta T_{UU}(\tilde{U}')} \{\sigma_{U_0}(U),T_{UU}(\tilde{U}')\}
\end{equation}
where $\tilde U_0$ is finite. Inserting \eqref{eq:sigTFinal} and manipulating the result yields
 \begin{equation}\label{eq:sigsig2}
 \begin{split}
    \{\sigma_{U_0}(U),\sigma_{\tilde U_0}(\tilde{U})\} = \frac{6\pi}{c}\bigg([ \Theta(&\tilde U-U)\Theta(\tilde U_0-U_0)-\Theta(U-\tilde U)\Theta(U_0-\tilde U_0)]\\
    &\times [\Theta(U-\tilde U_0)\Theta(\tilde U-U_0) + \Theta(\tilde U_0-U)\Theta(U_0-\tilde U)]\\
    + [\Theta&(\tilde U-\tilde U)\Theta(U_0-\tilde U_0)+\Theta(U-\tilde U)\Theta(\tilde U_0-U_0)]\\
    &\times [\Theta(U-\tilde U_0)\Theta(U_0-\tilde U)-\Theta(\tilde U_0-U)\Theta(\tilde U-U_0)] \\
    +(\tilde u& -u)\delta(U_0-\tilde U_0) + (u-u_0)(\tilde u-\tilde u_0)\delta'(U_0-\tilde U_0)\\
    -(\tilde u&-\tilde u_0)\delta(U-\tilde U_0) + (u-u_0)\delta(\tilde U -U_0) \bigg)
 \end{split}
\end{equation}
 with corresponding results for $\{\sigma_{V_0}(V),\sigma_{\tilde V_0}(\tilde V)\}$.  It turns out this form of the $\sigma$ commutator is somewhat cumbersome to use in our calculations below.  The calculations are simplified by making use of an equivalent relation in which antisymmetry is no longer manifest, but which is more compact:
 \begin{equation}\label{eq:sigsig3}
 \begin{split}
    \{\sigma_{U_0}(U),\sigma_{\tilde U_0}(\tilde{U})\} = \frac{6\pi}{c}\bigg(&\Theta(\tilde U-U)\Theta(U-\tilde U_0)-\Theta(U-\tilde U)\Theta(\tilde U_0-U) \\
    &- \Theta(\tilde U-U_0)\Theta(U_0-\tilde U_0) + \Theta(U_0-\tilde U)\Theta(\tilde U_0-U_0) \\
    &+(\tilde u -u)\delta(U_0-\tilde U_0) + (u-u_0)(\tilde u-\tilde u_0)\delta'(U_0-\tilde U_0)\\
    &-(\tilde u-\tilde u_0)\delta(U-\tilde U_0) + (u-u_0)\delta(\tilde U -U_0) \bigg).
 \end{split}
\end{equation}

%%%%%%%%%%%%%%%%%%%%%%%%%%%%%%%
\subsection{The HRT-area algebra}\label{sec:area}

We now we turn to the commutator of HRT-areas.
Using the results from Sections \ref{sec:sigT} and \ref{sec:sigsig}, we can compute the semiclassical commutator of an HRT-area operator with the boundary stress tensor, as well as that between two area operators $A_{HRT}(U_1,V_1,U_2,V_2)$ and $A_{HRT}(U_1',V_1',U_2',V_2')$.  As in section \ref{sec:HRTAact},  the arguments denote the coordinates of the two anchor points that define $\partial R$.  We again use the renormalized area operator \eqref{eq:renormA} whose dependence on $\sigma$ is given both by the explicit term in \eqref{eq:Anew} and the dependence of \eqref{eq:AHRTvacexp} on $\sigma$ through $U(u)$ and $V(v)$.  Using $\tilde A_{HRT}$ to denote the renormalized area in the conformal frame where the stress tensor vanishes (and where the boundary metric is $-dudv$), the second of these takes the explicit form
\begin{equation}
    \begin{split}
       \tilde A_{HRT}(U_1,V_1,U_2,V_2)
       =& \ln [ 4(u_1-u_2)(v_2-v_1)]  \\
       =& \ln \biggr[ 4 \int_{U_2}^{U_1}dU e^{-2\sigma_-(U)} \int_{V_1}^{V_2} dV e^{-2\sigma_+(V)} \biggr].
    \end{split}
\end{equation}
Taking functional derivatives with respect to $\sigma_-$ yields
\begin{equation}\label{eq:delAdels}
\begin{split}
    \frac{\delta A_{HRT}(U_1,V_1,U_2,V_2) }{\delta \sigma_-(U)} =& -\frac{2e^{-2\sigma_-(U)}}{\bigg|\int_{U_2}^{U_1}dU'' e^{-2\sigma_-(U'')}\bigg|}\Theta(\max(U_1,U_2)-U)\Theta(U-\min(U_1,U_2)) \\
    =& -\frac{2e^{-2\sigma_-(U)}}{|u(U_1)-u(U_2)|}\Theta(\max(U_1,U_2)-U)\Theta(U-\min(U_1,U_2)),
\end{split}
\end{equation}
with an analogous expression for the functional derivative of the area with respect to $\sigma_+$.
%%%%%%%%%%%%%%%%%%%%%%%%%%%%%%%%%%%%%%%%%%%%%
\subsubsection{The area operator and stress-energy tensor commutator}\label{sec:AT}
We will now use the above results to understand the commutator of an HRT area operator with the boundary stress tensor.  We introduce the notation $\tilde A_{HRT} \equiv \tilde A_{U_0,V_0}$, where $U_0$ and $V_0$ are the finite points at which $\sigma_{U_0}(U)$ and $\sigma_{V_0}(V)$ are independent of the stress tensor. From Eq.~\eqref{eq:delAdels}, we find
\begin{equation}\label{eq:ATvac}
    \begin{split}
        \{\tilde A_{U_0,V_0}(U_1,V_1,U_2,V_2),T_{UU}(U)\} =& \int_{U_2}^{U_1} dU' \frac{\delta \tilde A_{U_0,V_0}(U_1,V_1,U_2,V_2)}{\delta \sigma_{U_0}(U')} \{\sigma_{U_0}(U'),T_{UU}(U)\} \\
        =& -\frac{2}{u(U_1)-u(U_2)} \int_{U_2}^{U_1} dU'  e^{-2\sigma_{U_0}(U')} \{\sigma_{U_0}(U'),T_{UU}(U)\}.
    \end{split}
\end{equation}
Using Eq.~\eqref{eq:Anew}, we see the full commutator is given by
\begin{equation}
\label{eq:Aneedssig}
    \{A_{U_0,V_0}(U_1,V_1,U_2,V_2),T_{UU}(U)\} = \{\tilde A_{U_0,V_0},T_{UU}(U)\} + \{\sigma_{U_0}(U_1),T_{UU}(U)\} + \{\sigma_{U_0}(U_2),T_{UU}(U)\}.
\end{equation}
Using the explicit form of $\{\sigma_{U_0}(U'),T_{UU}(U)\}$ as given in Eq.~\eqref{eq:sigTFinal}, most of the resulting terms cancel among themselves after insertion into \eqref{eq:Aneedssig}.
The result reduces to
\begin{equation}\label{eq:ATphys}
    \begin{split}
        \{A_{HRT}(U_1,V_1,U_2,V_2),T_{UU}(U)\} =&  -\frac{2}{u(U_1)-u(U_2)} \int_{U_2}^{U_1} dU'  e^{-2\sigma_{U_0}(U')} \{\sigma_{U_0}(U'),T_{UU}(U)\}_{phys} \\
        &+ \{\sigma_{U_0}(U_1),T_{UU}(U)\}_{phys} + \{\sigma_{U_0}(U_2),T_{UU}(U)\}_{phys}.
    \end{split}
\end{equation}
where
\begin{equation}\label{eq:sigTphys}
    \{\sigma_{U_0}(U'),T_{UU}(U)\}_{phys} = \sigma_{U_0}'(U)\delta(U'-U)- \frac{1}{2} \delta'(U'-U).
\end{equation}
Plugging Eq.~\eqref{eq:sigTphys} into Eq.~\eqref{eq:ATphys} gives the final results
\begin{equation}\label{eq:AT}
    \begin{split}
        \biggr\{\frac{A_{HRT}(U_1,V_1,U_2,V_2)}{4G} ,T_{UU}(U)\biggr\} =& \frac{1}{4G[u(U_1)-u(U_2)]}[e^{-2\sigma_{U_0}(U_1)}\delta(U_1-U) - e^{-2\sigma_{U_0}(U_2)}\delta(U_2-U)] \\
        &+\frac{\sigma_{U_0}'(U)}{4G}\delta(U_1-U) - \frac{1}{8G} \delta'(U_1-U) \\
        &+\frac{\sigma_{U_0}'(U)}{4G}\delta(U_2-U) - \frac{1}{8G} \delta'(U_2-U), \\
        \biggr\{\frac{A_{HRT}(U_1,V_1,U_2,V_2)}{4G},T_{VV}(V)\biggr\} =& \frac{1}{4G[v(V_1)-v(V_2)]}[e^{-2\sigma_{V_0}(V_1)}\delta(V_1-V) - e^{-2\sigma_{V_0}(V_2)}\delta(V_2-V)] \\
        &+\frac{\sigma_{V_0}'(V)}{4G}\delta(V_1-V) - \frac{1}{8G} \delta'(V_1-V) \\
        &+\frac{\sigma_{V_0}'(V)}{4G}\delta(V_2-V) - \frac{1}{8G} \delta'(V_2-V).
    \end{split}
\end{equation}
While the right-hand side appears to depend on $U_0$, the $U_0$ dependence of $\sigma_{U_0}$ (and thus of $u(U)$) is determined by \eqref{eq:T} and the boundary conditions that both $\sigma_{U_0}(U)$ and its first $U$ derivative vanish at $U_0$.  Using this result, a careful calculation shows the right-hand side to be independent of $U_0$.

Note that for the special case $U_1=0=V_1$, $U_2 = - \infty$, $V_2=\infty$ with $\sigma_{U_0}(U) = \sigma_{V_0}(V) =0$, our \eqref{eq:AT} reduces to the $\lambda$-derivatives of \eqref{eq:AtransTijU} and \eqref{eq:AtransTijV} evaluated at $\lambda =0$.  This establishes the consistency of the above with the results of sections \ref{sec:kink} and \ref{sec:AdS3}.

%%%%%%%%%%%%%%%%%%%%%%%%%%%%%%%%%%%%%%%%%%
\subsubsection{The commutator between two area operators}\label{sec:areacomm}
Our final task will be to write the semiclassical commutator of two HRT area operators in a similar fashion. As in section \ref{sec:HRTAact}, we write any $A_{HRT}(U_1,V_1,U_2,V_2)$ in the form

\begin{equation}
\label{eq:AHRTlamvacexp2}
A_{HRT}(U_1,V_1,U_2,V_2) = A_{U}(U_1,U_2) + A_{V}(V_1,V_2) + 2 \ln 2,
\end{equation}
and similarly $\tilde A_{HRT}= \tilde A_{U}(U_1,U_2) + \tilde A_{V}(V_1,V_2) + 2 \ln 2$.  However, to make manifest the dependence of the renormalized HRT-area on $U_0$ and $V_0$ as functionals of the stress tensor, we will instead use the notation $\tilde A_U \equiv \tilde A_{U_0}$ and $\tilde A_V \equiv A_{V_0}$, where $U_0$ and $V_0$ are defined as above.  Noting that the $U$ parts are functionals of the right-moving stress tensor while the $V$ parts are functionals of the left-moving stress tensor, we see that the $U$ and $V$ parts commute with each other.
We may then focus on the commutator between two $U$ parts with the understanding that results for the $V$ commutators can be recovered using the symmetry $U \rightleftarrows V$.

For two HRT surfaces anchored respectively at $(U_1,U_2)$ and $(U_1',U_2')$, we have
\begin{equation}\label{eq:AAU}
    \begin{split}
        \{A_{U}(U_1,U_2),A_{U}(U_1', U_2')\} =&         \{\tilde A_{U_0}(U_1,U_2),\tilde A_{U_0'}(U_1', U_2')\} \\
        &+ \{\tilde A_{U_0}(U_1,U_2),\sigma_{U_0'}(\tilde{U}_1)\} \\
        &+ \{\tilde A_{U_0}(U_1,U_2),\sigma_{U_0'}(\tilde{U}_2)\} \\
        &+ \{\sigma_{U_0}(U_1),\tilde A_{U_0'}(U_1',U_2')\} \\
        &+
        \{\sigma_{U_0}(U_2),\tilde A_{U_0'}(U_1',U_2')\} \\
        &+ \{\sigma_{U_0}(U_1),\sigma_{U_0'}({U}_1')\} + \{\sigma_{U_0}(U_1),\sigma_{U_0'}({U}_2')\} \\
        &+ \{\sigma_{U_0}(U_2),\sigma_{U_0'}({U}_1')\} + \{\sigma_{U_0}(U_2),\sigma_{U_0'}({U}_2')\}.
    \end{split}
\end{equation}
The first term in the above expression is given by
\begin{equation}
\label{eq:twotilde}
    \begin{split}
         \{\tilde A_{U_0}(U_1,U_2),\tilde A_{U_0'}(U_1', U_2')\}  = \\
        \int_{U_2}^{U_1}dU\int_{{U}_2'}^{{U}_1'} d{U'}& \frac{\delta \tilde A_{U_0}(U_1,U_2)}{\delta \sigma_{U_0}(U)}  \frac{\delta \tilde A_{U_0'}(U_1',U_2')}{\delta \sigma_{U_0'}({U'})}\{\sigma_{U_0}(U),\sigma_{U_0'}({U'})\},
    \end{split}
\end{equation}
and the next four terms will have forms analogous to
\begin{equation}
\label{eq:onetilde}
    \begin{split}
         \{\tilde A_{U_0}(U_1,U_2),\sigma_{U_0'}({U'})\} = \int_{U_2}^{U_1}dU \frac{\delta \tilde A_{U_0}(U_1,U_2)}{\delta \sigma_{U_0}(U)}\{\sigma_{U_0}(U),\sigma_{U_0'}({U'})\}.
    \end{split}
\end{equation}

We can evaluate \eqref{eq:twotilde} and \eqref{eq:onetilde} using \eqref{eq:sigsig3} and \eqref{eq:delAdels}. As in Section \ref{sec:AT}, when we insert the explicit form of $\{\sigma_{U_0}(U),\sigma_{U_0'}(U')\}$ as given by Eq.~\eqref{eq:sigsig3} into either \eqref{eq:twotilde} \eqref{eq:onetilde}, the majority of the resulting terms cancel among themselves.  In each case, the remaining terms can be obtained using a simplified $\sigma$ commutator given by only the first term in Eq.~\eqref{eq:sigsig3}, i.e., by using
\begin{equation}\label{eq:sigsigphys}
    \{\sigma_{U_0}(U),\sigma_{\tilde U_0}(\tilde U)\}_{truncated} = \frac{6\pi}{c}[\Theta(\tilde U-U)\Theta(U-\tilde U_0)-\Theta(U-\tilde U)\Theta(\tilde U_0-U)].
\end{equation}
We are now ready to insert this truncated $\sigma$ commutator into Eq.~\eqref{eq:AAU} to find the full commutator.\footnote{Although Eq.~\ref{eq:sigsigphys} has some remaining $\tilde U_0$-dependence, the final result will be manifestly independent of both $\tilde U_0$ and $U_0$.  This is reassuring, as $U_0$ and $\tilde U_0$ were arbitrary parameters that appeared in writing the HRT-areas as functionals of the stress tensor}

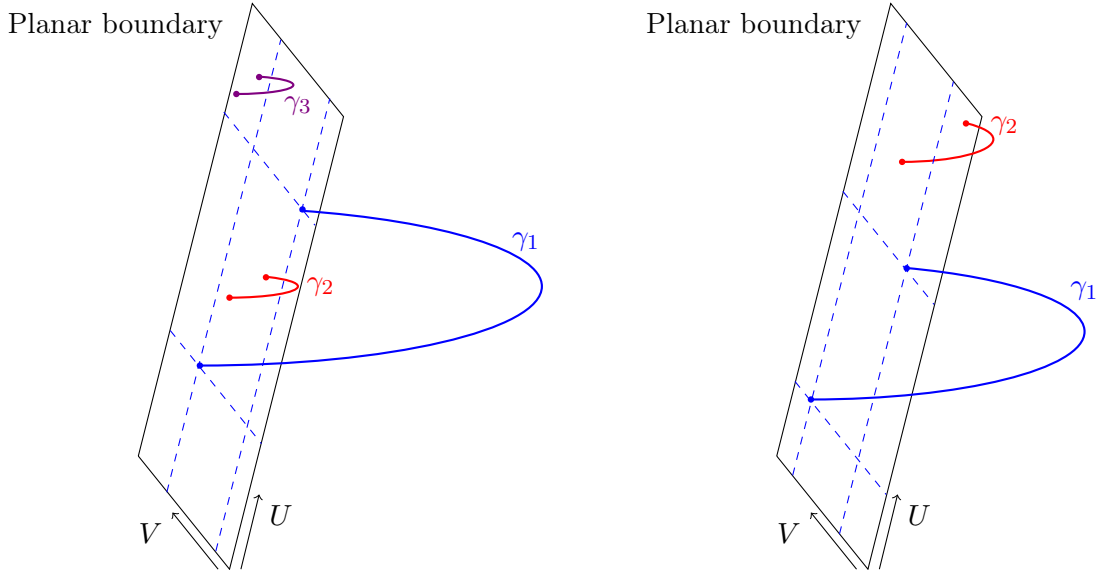
\begin{figure}
    \centering
    \begin{tikzpicture}[scale=3]
    \draw (-1,-1) -- (-.5,1) -- (-.1,.5) -- (-.6,-1.5) -- (-1,-1);
    \draw[red, thick] (-.6,-.3) arc[start angle=-90,delta angle=145,x radius=3mm, y radius=.5mm];
    \draw[red] (-.2,-.25) node {$\gamma_2$};
    \fill[red] (-.6,-.3) circle (.4pt);
    \fill[red] (-.44,-.21) circle (.4pt);
    \draw[blue, thick] (-.73,-.6) arc[start angle=-90,delta angle=163,x radius=15mm, y radius=3.5mm];
    \draw[blue] (.7,-.05) node {$\gamma_1$};
    \fill[blue] (-.73,-.6) circle (.4pt);
    \fill[blue] (-.28,.09) circle (.4pt);
    \draw (-1.1,.9) node {Planar boundary};
    \draw[->] (-.55,-1.5) -- (-.47,-1.17) node [below right] {$U$};
    \draw[->] (-.65,-1.5) -- (-.85,-1.25) node [below left] {$V$};
    \draw[dashed, blue] (-.875,-1.16) -- (-.375,.84);
    \draw[dashed, blue] (-.66,-1.42) -- (-.16,.58);
    \draw[dashed, blue] (-.86,-.445) -- (-.46,-.945);
    \draw[dashed, blue] (-.625,.52) -- (-.225,.02);
    \draw[violet, thick] (-.57,.6) arc[start angle=-90,delta angle=155,x radius=2.5mm, y radius=.4mm];
    \fill[violet] (-.57,.6) circle (.4pt);
    \fill[violet] (-.47,.675) circle (.4pt);
    \draw[violet] (-.3,.55) node {$\gamma_3$};

\begin{scope}[xshift=2.8cm]
    \draw (-1,-1) -- (-.5,1) -- (-.1,.5) -- (-.6,-1.5) -- (-1,-1);
    \draw[red, thick] (-.45,.3) arc[start angle=-90,delta angle=135,x radius=4mm, y radius=1mm];
    \draw[red] (0,.46) node {$\gamma_2$};
    \fill[red] (-.45,.3) circle (.4pt);
    \fill[red] (-.17,.47) circle (.4pt);
    \draw[blue, thick] (-.85,-.75) arc[start angle=-90,delta angle=160,x radius=12mm, y radius=3mm];
    \draw[blue] (.35,-.27) node {$\gamma_1$};
    \fill[blue] (-.85,-.75) circle (.4pt);
    \fill[blue] (-.43,-.17) circle (.4pt);
    \draw (-1.1,.9) node {Planar boundary};
    \draw[->] (-.55,-1.5) -- (-.47,-1.17) node [below right] {$U$};
    \draw[->] (-.65,-1.5) -- (-.85,-1.25) node [below left] {$V$};
    \draw[dashed, blue] (-.93,-1.085) -- (-.43,.915);
    \draw[dashed, blue] (-.725,-1.345) -- (-.225,.655);
    \draw[dashed, blue] (-.92,-.67) -- (-.52,-1.17);
    \draw[dashed, blue] (-.71,.17) -- (-.31,-.33);
\end{scope}
    
\end{tikzpicture}
    \caption{ Various possible relative configurations for the anchor points of HRT surfaces. In the left panel, the $U$ coordinates of the anchors of $\gamma_2$ define an interval that is fully contained in the corresponding $U$ interval for $\gamma_1$, while the anchors of $\gamma_3$ define a $U$ interval that does not intersect that defined by $\gamma_1$.     In such cases the $U$ parts of the HRT-areas commute.  The same statements hold with $U$ replaced by $V$.  In contrast, in the right panel the $U$ intervals defined by $\gamma_1$ and $\gamma_2$ intersect without one being fully contained in the other. In this case the commutator of the $U$ parts of the HRT-areas will not vanish.}
    \label{fig:area_algebra}
\end{figure}

Due to the step functions in \eqref{eq:sigsigphys}, it is convenient to divide the calculation into cases.  Let us first consider the cases shown at left in figure \ref{fig:area_algebra}, where the intervals $(U_1,U_2)$ and $(U_1',U_2')$ either have no intersection or  where one interval is fully contained in the other.  These two situations are equivalent due to the symmetry under interchange of $R$ with $\bar R$.  For this case, one finds that the various terms cancel to give
\begin{equation}
    \{A_U(U_1,U_2),A_U(U_1',U_2')\}_{U} = 0.
\end{equation}
This should be no surprise as, depending on the $V$-values of the anchor points, this case allows the two HRT surfaces to be spacelike separated.

% \begin{figure}
%     \centering
%     \includegraphics[scale=.66]{nonoverlap.png}
%     \caption{The two possibilities for non-overlapping HRT surfaces $\gamma_1$ and $\gamma_2$. This also includes the case when they meet at an anchor point. The $u$-axis is shown the black line along the boundary, while the $v$-axis runs perpendicularly out of the page.}
%     \label{fig:nonoverlap}
% \end{figure}

% \begin{figure}
%     \centering
%     \includegraphics[scale=.9]{overlap.png}
%     \caption{Two overlapping HRT surfaces, $\gamma_1$ and $\gamma_2$.}
%     \label{fig:overlap}
% \end{figure}

The remaining case occurs when the intervals $(U_1,U_2)$ and $(U_1',U_2')$ overlap without having one fully contained in the other; see right panel of figure \ref{fig:area_algebra}.  For notational simplicity we let $u_i=u(U_i), u_i'=u(U_i')$ and we take both $U_1>U_2$ and $U_1'>U_2'$. In this case we find
\begin{equation}\label{eq:AA}
    \{A_U(U_1,U_2),A_U(U_1',U_2')\} =
        \begin{cases} -\frac{6\pi}{c}\biggr( 1 - \frac{2(u_1-u'_1)(u_2-u'_2)}{(u_1-u_2)(u'_1-u'_2)} \biggr), & U_2' <U_2 < U'_1 < U_1 \\
        \frac{6\pi}{c}\biggr( 1 - \frac{2(u_1-u'_1)(u_2-u'_2)}{(u_1-u_2)(u'_1-u'_2)} \biggr), & U_2 < U'_2 <U_1 < U'_1,\\
        \end{cases}
\end{equation}
again with analogous results for the $V$ parts. We see there is no remaining dependence on $U_0$ or $U_0'$, 

Let us now further explore this result by studying special cases.  We begin by noting that choosing $U_1=0$, $V_1=0$ and $U_2=-\infty$, $V_2=\infty$ sets $A_{HRT}(U_1,V_1,U_2,V_2)=A_{HRT}[R_0]$, which is the case studied previously\footnote{Section \ref{sec:AdS3}
in fact defined $R_0$ only in the Poincar\'e vacuum, but here we generalize the definition so that in any spacetime $R_0$ is the region between $(U,V)=(0,0)$ and $(U,V)= (-\infty, \infty)$.} in section \ref{sec:AdS3}.   Combining expression \eqref{eq:AA} with the corresponding result for $V$ then yields
\begin{equation}
\{A_{HRT}[R_0],A_{HRT}(U'_1,V'_1,U'_2,V'_2))\} = -8\pi G \left( \frac{u'_1 \Theta(-u'_1 u'_2)}{u'_1 - u'_2} + \frac{v'_1 \Theta(-v'_1 v'_2)}{v'_2 - v'_1}\right).
\end{equation}
To evaluate this on the AdS$_3$ vacuum we set $u'_i=U'_i, v'_i=V'_i$.  The result then agrees with \eqref{eq:AAgeo2}.   We thus conclude that \eqref{eq:AA} is the generalization of \eqref{eq:AAgeo} to general intervals and to arbitrary solutions in our phase space.

Another interesting special case arises where we again evaluate the commutator on the AdS$_3$ vacuum (thus setting $u_i=U_i, u'_i=U'_i$), but where we take all of the anchor points to lie on a $t=constant$ slice of the boundary.  Choosing this slice to be $t=0$, this is equivalent to setting $u_i = U_i = - V_i = -v_i$, $u_i' = U_i' = - V_i'=-v_i'$.  As a result, if the $U$ term gives the upper result on the right-hand-side of \eqref{eq:AA}, then the $V$ term gives the analogue of the lower result and the two cancel; i.e., when $R_1, R_2$ are both subsets of the $t=0$ slice on the boundary we find

\begin{equation}
    \{A_{HRT}[R_1],A_{HRT}[R_2]\} = 0.
\end{equation}
This result is to be expected from the fact that commutators of real functions must change sign under time reversal (as indicated in the quantum mechanical context due to the required factor of $i$ in the commutator), while the specified configuration and background are manifestly invariant under time-reversal.  Indeed, for this reason the leading semiclassical commutator of HRT-areas will always vanish on a background that enjoys a time-reversal symmetry that leaves invariant both $R_1$ and $R_2$.

%%%%%%%%%%%%%%%%%%%%%%%%%%%%%%%%%%%%%%%%%%%%%%%%%%%%%%%%%%%%%%%%%%%%%%%%%%%%%%%%%%%

\section{Discussion}
\label{sec:disc}

The goal of our work was to study the flow on phase space generated by HRT-areas $A_{HRT}[R]$ in Einstein-Hilbert gravity, filling in various gaps in the literature. In particular, we showed that the canonical commutation relations can be used to evaluate this flow on any bulk Cauchy surface $\Sigma$ passing through the HRT surface $\gamma_R$.  On such surfaces, the flow leaves the induced metric invariant but shifts the extrinsic curvature by a delta-function as described by \eqref{eq:AKPB}.  As predicted in \cite{Bousso:2020yxi}, this effectively boosts the entanglement wedge of $R$ relative to that of the complementary region $\bar R$.  However, the effect on the region to the future or past of $\gamma_R$ must be determined by solving the bulk equations of motion in the presence of appropriate boundary conditions.

Such boundary conditions lead to a difference between HRT-area flow and the kink transformation of \cite{Bousso:2020yxi}.  This difference was again predicted in \cite{Bousso:2020yxi}.  For vacuum AdS$_3$ spacetimes one can compute this difference and use it to obtain explicit formulae for the action of our flow on both the boundary stress tensor and other HRT-areas.  Results were presented in section \ref{sec:AdS3} for quantities evaluated on the Poincar\'e AdS$_3$ vacuum.  Section \ref{sec:conf} then used a different approach to evaluate the associated commutators on general vacuum solutions asymptotic to AdS$_3$.  This latter approach was based on the fact that, since all such solutions can be generated from the Poincar\'e vacuum by acting with boundary conformal transformations, any observable in this context can in principle be written as a functional of the boundary stress tensor.  This method may also be of interest in its own right for computing other commutators for which the action on initial data is more complicated than that of $A_{HRT}[R]$.

We note that \cite{Bousso:2019dxk} also studied the effect of applying the transformation \eqref{eq:AKPB} at {\it non}-extremal codimension-2 spacelike surfaces $\gamma$.  Again, this corresponds to boosting the initial data in what one might call the right wedge relative to that in the complementary (left) wedge.  It was noted in \cite{Bousso:2019dxk} that when $\gamma$ is non-extremal the resulting initial data fails to satisfy the constraint equations of Einstein-Hilbert gravity.  From our perspective, this is no surprise.  For extremal $\gamma$ our \eqref{eq:AKPB} is generated by a diffeomorphism invariant observable, which necessarily commutes with all constraints.  But more generally we would expect this to fail.  In particular, while the flow generated by any diffeomorphism-invariant $A[\gamma]$ must also preserve the constraints, for non-extremal $\gamma$ there will be a non-trivial contribution from the first term in \eqref{eq:chain1}, so that the flow would no longer be given simply by \eqref{eq:AKPB}.  In this case the contribution from the first term in \eqref{eq:chain1} must precisely cancel the contribution from the constraint-violating part of \eqref{eq:AKPB}.

Some of our results may have further implications for holography, especially in connection with tensor network models of quantum error correction \cite{Pastawski:2015qua,Hayden:2016cfa}.  One such result is that the commutator of two HRT-areas vanishes at leading semiclassical order when evaluated on a background where both HRT-surfaces lie in a common surface of time-symmetry.\footnote{This in fact follows directly from time symmetry and did not require detailed computation.}  As a result, such HRT-areas may be specified simultaneously with high accuracy.  This observation may be useful in constructing bulk analogues of the above tensor networks (e.g. as in \cite{Bao:2018pvs,Bao:2019fpq}) which appears to require bulk states in which such areas are sharply peaked \cite{Akers:2018fow,Dong:2018seb}.

Another result that deserves further investigation was the observation in section \ref{sec:AdS3Tijact} that HRT-area flow produces states of infinite energy.  As argued in \cite{Ceyhan:2018zfg,Faulkner:2018faa,Bousso:2019dxk}, the development of a UV singularity should be no surprise.  But note that any sharp quantum eigenstate of $A_{HRT}[R]$ will be invariant under the flow that this operator generates.    In particular, the expectation value of the energy will not change under this flow.  We thus conclude that the expected energy in such states must be divergent, or at least set by some UV regulator.  This may again have implications for the use of tensor networks in holography.  More generally, this feature may be relevant for understanding the sense in which holographic quantum error correcting codes decompose into superselection sectors defined by $A_{HRT}[R]$ \cite{Harlow:2016vwg}.  The point here is that the relevant code subspace is often taken to be states of low energy in the dual gauge theory \cite{Almheiri:2014lwa}, while we now see explicitly that states in any given such superselection sector must have energies set by the UV cutoff.  However, as we see from the 2+1 dimensional case, this need not cause large curvatures in the bulk.

With regard to future directions, we recall that our work focussed on Einstein-Hilbert gravity.  But it is natural to expect similar results to hold in the presence of higher derivative corrections.  This will be explored in the forthcoming work \cite{DMRGFLOW}, which will also comment further on the relation of $A_{HRT}[R]$ to the modular Hamiltonians on $R$ and $\bar R$.

\acknowledgments

It is a pleasure to thank Xi Dong and Pratik Rath for many conversations related to this work.  This material is based upon work supported by the Air Force Office of Scientific Research under award number FA9550-19-1-0360, and by funds from the University of California.

\appendix

\section{Normalizations and the one-sided boost}
\label{app:norm}

As discussed in section \ref{sec:kinkrel}, both the flow generated by $A_{HRT}$ and the kink transformation can be described as a sort of one-sided boost.  This appendix verifies the details of this relationship, and shows in particular that such a transformation leads precisely to \eqref{eq:AKPB} with the stated normalizations.  We perform an explicit computation below for spacetimes that admit a bulk Killing field $\xi^a$ which acts locally like a boost near $\gamma$.  We also study the effect on initial data defined for particularly convenient Cauchy surfaces.  But since the flow clearly does not affect initial data within either wedge, the delta-function terms in \eqref{eq:AKPB} can depend only on the local structure near $\gamma$.  It will thus be clear that the same normalizations continue to hold for the more general transformation described in section \ref{sec:kinkrel} (and originally defined in \cite{Bousso:2020yxi}),
where we allow arbitrary Cauchy surfaces through $\gamma$ and
where the boost operator is defined only in the local approximation where one replaces each plane orthogonal to $\gamma$ with flat Minkowski space.

We will call our one-sided boost $\eta_\alpha$.
Before proceeding, recall that $\gamma$ defines two entanglement wedges, one on each side of the surface, and that $\xi^a$ is past-directed in one (which we call the left wedge) and is future directed in the other (which we call the right wedge).
In the left entanglement we take $\eta_\alpha$ to be the identity. In the right wedge we instead take $\eta_\alpha$ to be the diffeomorphism generated by moving points along the orbits of the KVF $\xi^a$ by a Killing parameter $\alpha/\kappa$, where $\kappa$ is the surface gravity of $\xi^a$.     Thus $\eta_\alpha$ is a diffeomorphism in each entanglement wedge, though it is not smooth at $\gamma$ and we have not defined its action at points that lie inside\footnote{On the boundary of the future or past of $\gamma$ the action of $\eta_\alpha$ can be defined by continuity.} the chronological future or past of $\gamma$.

We thus examine the action of $\eta_\alpha$ on Cauchy surfaces that contain $\gamma$.
Note that $\eta_\alpha$ will map any such Cauchy surface $\Sigma$ to another such surface $\tilde \Sigma$. To simplify our discussion, we will make a special choice for the Cauchy surface $\Sigma$, or at least for the part of that Cauchy surface near $\gamma$.  We begin by choosing a unit spacelike vector field $m^a$ normal to $\gamma$ and defined smoothly everywhere on $\gamma$.  We then extend $m^a$ to some region near $\gamma$ by taking $m^a$ to be the unit affinely-parameterized tangent to a congruence of geodesics.  At least near $\gamma$, these geodesics will generate a hypersurface.  We take this to coincide with the part of $\Sigma$ near our HRT surface.\footnote{Had we started with a generic hypersurface, we could consider the family of spacetime geodesics which happen to be tangent to the hypersurface at $\gamma$.  The analysis would then be identical to leading order near $\gamma$, and in particular would give precisely the same delta-function terms in \eqref{eq:AKPB}.}  The normal $n^a$ to $\Sigma$ then satisfies $m^a n_a=0$ in the region near $\gamma$.   We then further extend both $m^a$ and $n^a$ off of the original slice $\Sigma$ in an arbitrary smooth manner that preserves the conditions $n^a m_a=0$ and $n_a n^a =-1$. Thus we have the useful relations
\begin{eqnarray}
\label{eq:extensions}
m^a \nabla_a  m^b &=& 0  \\
m^a n_b \nabla_a  n^b   &=&\frac{1}{2} m^a \nabla_a  (-1) = 0 \\
m^a m_b \nabla_a  n^b   &=& m^a \nabla_a  (m^b n_b) = 0.
\end{eqnarray}
We also use $s$ to denote proper distance along geodesics in the original congruence, with $s=0$ at $\gamma$.

Let us now construct a second Cauchy surface $\tilde \Sigma$ by applying a map $\eta_\alpha$ to $\Sigma.$ The map $\eta_\alpha$ is almost a diffeomorphism, except that it is not smooth at $\gamma.$  Before proceeding, note that $\gamma$ defines two entanglement wedges, one on each side of the surface, and that $\xi^a$ is past-directed in one (which we call the left wedge and in which we take $s<0$) and is future directed in the other (which we call the right wedge and in which we take $s>0$).
In the left entanglement we take $\eta_\alpha$ to be the identity. In the right wedge we instead take $\eta_\alpha$ to be the diffeomorphism generated by moving points along the orbits of the KVF $\xi^a$ by a Killing parameter $\alpha/\kappa$, where $\kappa$ is the surface gravity of $\xi^a$.     As a result, the normal $\tilde n^a$ to $\tilde \Sigma$ satisfies
\begin{equation}
\label{eq:relatenormalstoSigma}
\tilde n^a =  \cosh(\alpha \Theta(s)) n^a + \sinh(\alpha \Theta(s)) m^a
\dots,
\end{equation}
where $\dots$ denotes terms that are smooth but which depend on the way that our vector fields were extended off of the original Cauchy surface $\Sigma$. Similarly, the normal $\tilde m$ to $\gamma$ in $\tilde \Sigma$ satisfies
\begin{equation}
\tilde m^a =  \cosh(\alpha \Theta(\tilde s)) m^a + \sinh(\alpha \Theta(\tilde s)) n^a
\dots,
\end{equation}
where we have used the fact that $\tilde \Sigma$ is isometric to $\Sigma$ to introduce a coordinate $\tilde s$ that measures proper distance from $\gamma$ along $\tilde \Sigma$ in the same way that $s$ does along $\Sigma$.  Note that the conditions $m_a m^a = - n_a  n^a = 1$ and $n_a m^a=0$ give
\begin{eqnarray}
\label{eq:textensions}
\tilde m_b \nabla_a  n^b &=&0 \\
\tilde m_b \nabla_a  m^b &=&0  \nonumber .
\end{eqnarray}

Let us now recall that the extrinsic curvature of $\Sigma$ can be described by a degenerate tensor $K_{ab} = - h_a^c \nabla_c  n_b$ whose indices range over all coordinates of our spacetime (and not just those on $\Sigma$).  Although it is not manifest from the definition, this tensor is symmetric.  From \eqref{eq:relatenormalstoSigma}, we thus see that the extrinsic curvatures $K_{ab}$ and $\tilde K_{ab}$ of $\Sigma$ and $\tilde \Sigma$ are related everywhere by the action of $\eta_\alpha$ except for the component $\tilde m^a \tilde m^b \tilde K_{ab}$ which will be sensitive to derivatives of the theta-functions in \eqref{eq:relatenormalstoSigma}:
\begin{equation}
\label{eq:newK}
\tilde m^a \tilde m^b \tilde K_{ab} =  \tilde m^a \tilde m^b  \nabla_a  \tilde n_b.
\end{equation}
Such derivatives introduce $\delta$-function terms in $\tilde K_{ab}$ that are not present in the image\footnote{Since $\tilde K_{ab}$ is defined only on $\tilde \Sigma$, there is no need to define this flow in the causal past or future of $\gamma$.  Furthermore, since the extrinsic curvature $K_{ab}$ of $\Sigma$ is smooth at $\gamma$, we can define $\eta_\alpha^* K_{ab}$ at $\gamma$ by requiring it to be a smooth tensor on $\tilde \Sigma$ when expressed in terms of coordinates on $\tilde \Sigma$ obtained by acting with $\eta_\alpha$ on smooth coordinates for $\Sigma$.  We emphasize that $\tilde \Sigma$ can be regarded as an intrinsically-smooth manifold whose embedding in the bulk happens not to be smooth.} $\eta_\alpha^* K_{ab}$ of $K_{ab}$ under the flow $\eta_\alpha$.

Furthermore, since when acting on scalars we have $\tilde m^a \nabla_a = \partial_{\tilde s}$, we may write
\begin{equation}
\label{eq:dterms}
\tilde m^a  \nabla_a  \tilde n_b =
 \alpha \delta(\tilde s)  \tilde m_b + \left( \sinh(\alpha \Theta(\tilde  s))
\tilde m^a \tilde m^b  \nabla_a n_b + \cosh(\alpha \Theta(\tilde  s))
\tilde m^a \tilde m^b  \nabla_a m_b + \dots \right),
\end{equation}
where the final step uses \eqref{eq:textensions}.
Since the final $\dots$ terms in \eqref{eq:dterms} are smooth, they are determined by their values in the left and right wedges and must thus be a part of $\eta_\alpha^*  K_{ab}.$  We therefore conclude that the extrinsic curvatures of $\Sigma$ and $\tilde \Sigma$ are related by
\begin{equation}
\label{eq:newK2}
\tilde K_{ab} = \eta_\alpha^*  K_{ab} + \alpha \delta (\tilde s) \ \tilde m^a \tilde m^b,
\end{equation}
which agrees with \eqref{eq:AKPB} if we set $\alpha =-2\pi$ and take the bracket with $A_{HRT}/4G$ to give $d/d\lambda$.

As mentioned in footnote \ref{foot:norm}, the form of the normalization factor in \eqref{eq:newK2} differs from that presented in \cite{Bousso:2020yxi}.  This difference arises from the fact that the results of \cite{Bousso:2020yxi} were expressed using coordinates that are not smooth on $\tilde \Sigma$, and thus which introduce significant dependence on regulators.  In contrast, even though it is not smoothly embedded in the bulk, we emphasize that $\tilde \Sigma$ has the intrinsic structure of a smooth manifold, so that the corresponding proper distance coordinate $\tilde s$ is smooth on $\tilde \Sigma$.  This turns out to remove detailed dependence on regulators found in \cite{Bousso:2020yxi} and leads to the elegant result \eqref{eq:newK2}.

\bibliographystyle{jhep}
	\cleardoublepage

\renewcommand*{\bibname}{References}

\bibliography{algebra}

\providecommand{\href}[2]{#2}\begingroup\raggedright\begin{thebibliography}{10}

\bibitem{Ryu:2006bv}
S.~Ryu and T.~Takayanagi, {\it {Holographic derivation of entanglement entropy
  from AdS/CFT}},  {\em Phys. Rev. Lett.} {\bf 96} (2006) 181602,
  [\href{http://arxiv.org/abs/hep-th/0603001}{{\tt hep-th/0603001}}].

\bibitem{Ryu:2006ef}
S.~Ryu and T.~Takayanagi, {\it {Aspects of Holographic Entanglement Entropy}},
  {\em JHEP} {\bf 08} (2006) 045,
  [\href{http://arxiv.org/abs/hep-th/0605073}{{\tt hep-th/0605073}}].

\bibitem{Hubeny:2007xt}
V.~E. Hubeny, M.~Rangamani, and T.~Takayanagi, {\it {A Covariant holographic
  entanglement entropy proposal}},  {\em JHEP} {\bf 07} (2007) 062,
  [\href{http://arxiv.org/abs/0705.0016}{{\tt arXiv:0705.0016}}].

\bibitem{Peierls:1952cb}
R.~E. Peierls, {\it {The Commutation laws of relativistic field theory}},  {\em
  Proc. Roy. Soc. Lond. A} {\bf 214} (1952) 143--157.

\bibitem{Carlip:1993sa}
S.~Carlip and C.~Teitelboim, {\it {The Off-shell black hole}},  {\em Class.
  Quant. Grav.} {\bf 12} (1995) 1699--1704,
  [\href{http://arxiv.org/abs/gr-qc/9312002}{{\tt gr-qc/9312002}}].

\bibitem{Thiemann:1992jj}
T.~Thiemann and H.~A. Kastrup, {\it {Canonical quantization of spherically
  symmetric gravity in Ashtekar's selfdual representation}},  {\em Nucl. Phys.
  B} {\bf 399} (1993) 211--258, [\href{http://arxiv.org/abs/gr-qc/9310012}{{\tt
  gr-qc/9310012}}].

\bibitem{Kastrup:1993br}
H.~A. Kastrup and T.~Thiemann, {\it {Spherically symmetric gravity as a
  completely integrable system}},  {\em Nucl. Phys. B} {\bf 425} (1994)
  665--686, [\href{http://arxiv.org/abs/gr-qc/9401032}{{\tt gr-qc/9401032}}].

\bibitem{Kuchar:1994zk}
K.~V. Kuchar, {\it {Geometrodynamics of Schwarzschild black holes}},  {\em
  Phys. Rev. D} {\bf 50} (1994) 3961--3981,
  [\href{http://arxiv.org/abs/gr-qc/9403003}{{\tt gr-qc/9403003}}].

\bibitem{Jafferis:2014lza}
D.~L. Jafferis and S.~J. Suh, {\it {The Gravity Duals of Modular
  Hamiltonians}},  {\em JHEP} {\bf 09} (2016) 068,
  [\href{http://arxiv.org/abs/1412.8465}{{\tt arXiv:1412.8465}}].

\bibitem{Ceyhan:2018zfg}
F.~Ceyhan and T.~Faulkner, {\it {Recovering the QNEC from the ANEC}},  {\em
  Commun. Math. Phys.} {\bf 377} (2020), no.~2 999--1045,
  [\href{http://arxiv.org/abs/1812.04683}{{\tt arXiv:1812.04683}}].

\bibitem{Faulkner:2018faa}
T.~Faulkner, M.~Li, and H.~Wang, {\it {A modular toolkit for bulk
  reconstruction}},  {\em JHEP} {\bf 04} (2019) 119,
  [\href{http://arxiv.org/abs/1806.10560}{{\tt arXiv:1806.10560}}].

\bibitem{Bousso:2019dxk}
R.~Bousso, V.~Chandrasekaran, and A.~Shahbazi-Moghaddam, {\it {From black hole
  entropy to energy-minimizing states in QFT}},  {\em Phys. Rev. D} {\bf 101}
  (2020), no.~4 046001, [\href{http://arxiv.org/abs/1906.05299}{{\tt
  arXiv:1906.05299}}].

\bibitem{Bousso:2020yxi}
R.~Bousso, V.~Chandrasekaran, P.~Rath, and A.~Shahbazi-Moghaddam, {\it {Gravity
  dual of Connes cocycle flow}},  {\em Phys. Rev. D} {\bf 102} (2020), no.~6
  066008, [\href{http://arxiv.org/abs/2007.00230}{{\tt arXiv:2007.00230}}].

\bibitem{Lewkowycz:2018sgn}
A.~Lewkowycz and O.~Parrikar, {\it {The holographic shape of entanglement and
  Einstein\textquoteright{}s equations}},  {\em JHEP} {\bf 05} (2018) 147,
  [\href{http://arxiv.org/abs/1802.10103}{{\tt arXiv:1802.10103}}].

\bibitem{Chen:2018rgz}
Y.~Chen, X.~Dong, A.~Lewkowycz, and X.-L. Qi, {\it {Modular Flow as a
  Disentangler}},  {\em JHEP} {\bf 12} (2018) 083,
  [\href{http://arxiv.org/abs/1806.09622}{{\tt arXiv:1806.09622}}].

\bibitem{Jafferis:2015del}
D.~L. Jafferis, A.~Lewkowycz, J.~Maldacena, and S.~J. Suh, {\it {Relative
  entropy equals bulk relative entropy}},  {\em JHEP} {\bf 06} (2016) 004,
  [\href{http://arxiv.org/abs/1512.06431}{{\tt arXiv:1512.06431}}].

\bibitem{Donnelly:2016auv}
W.~Donnelly and L.~Freidel, {\it {Local subsystems in gauge theory and
  gravity}},  {\em JHEP} {\bf 09} (2016) 102,
  [\href{http://arxiv.org/abs/1601.04744}{{\tt arXiv:1601.04744}}].

\bibitem{Speranza:2017gxd}
A.~J. Speranza, {\it {Local phase space and edge modes for
  diffeomorphism-invariant theories}},  {\em JHEP} {\bf 02} (2018) 021,
  [\href{http://arxiv.org/abs/1706.05061}{{\tt arXiv:1706.05061}}].

\bibitem{Chandrasekaran:2019ewn}
V.~Chandrasekaran and K.~Prabhu, {\it {Symmetries, charges and conservation
  laws at causal diamonds in general relativity}},  {\em JHEP} {\bf 10} (2019)
  229, [\href{http://arxiv.org/abs/1908.00017}{{\tt arXiv:1908.00017}}].

\bibitem{Headrick:2007km}
M.~Headrick and T.~Takayanagi, {\it {A Holographic proof of the strong
  subadditivity of entanglement entropy}},  {\em Phys. Rev. D} {\bf 76} (2007)
  106013, [\href{http://arxiv.org/abs/0704.3719}{{\tt arXiv:0704.3719}}].

\bibitem{Marolf:2010tg}
D.~Marolf, M.~Rangamani, and M.~Van~Raamsdonk, {\it {Holographic models of de
  Sitter QFTs}},  {\em Class. Quant. Grav.} {\bf 28} (2011) 105015,
  [\href{http://arxiv.org/abs/1007.3996}{{\tt arXiv:1007.3996}}].

\bibitem{Casini:2011kv}
H.~Casini, M.~Huerta, and R.~C. Myers, {\it {Towards a derivation of
  holographic entanglement entropy}},  {\em JHEP} {\bf 05} (2011) 036,
  [\href{http://arxiv.org/abs/1102.0440}{{\tt arXiv:1102.0440}}].

\bibitem{Lewkowycz:2013nqa}
A.~Lewkowycz and J.~Maldacena, {\it {Generalized gravitational entropy}},  {\em
  JHEP} {\bf 08} (2013) 090, [\href{http://arxiv.org/abs/1304.4926}{{\tt
  arXiv:1304.4926}}].

\bibitem{DMRGFLOW}
X.~Dong, D.~Marolf, and P.~Rath, ``{Geometric entropies and their geometric
  flow: the power of Lorentzian methods}.'' {to appear}.

\bibitem{DMRJLMS}
X.~Dong, D.~Marolf, and P.~Rath, ``{The JLMS Formula, Modular Flow and the Area
  Operator}.'' {to appear}.

\bibitem{Engelhardt:2018kcs}
N.~Engelhardt and A.~C. Wall, {\it {Coarse Graining Holographic Black Holes}},
  {\em JHEP} {\bf 05} (2019) 160, [\href{http://arxiv.org/abs/1806.01281}{{\tt
  arXiv:1806.01281}}].

\bibitem{DiFrancesco:1997nk}
P.~Di~Francesco, P.~Mathieu, and D.~Senechal, {\em {Conformal Field Theory}}.
\newblock Graduate Texts in Contemporary Physics. Springer-Verlag, New York,
  1997.

\bibitem{Morrison_2013}
I.~A. Morrison and M.~M. Roberts, {\it Mutual information between thermo-field
  doubles and disconnected holographic boundaries},  {\em Journal of High
  Energy Physics} {\bf 2013} (Jul, 2013).

\bibitem{Pastawski:2015qua}
F.~Pastawski, B.~Yoshida, D.~Harlow, and J.~Preskill, {\it {Holographic quantum
  error-correcting codes: Toy models for the bulk/boundary correspondence}},
  {\em JHEP} {\bf 06} (2015) 149, [\href{http://arxiv.org/abs/1503.06237}{{\tt
  arXiv:1503.06237}}].

\bibitem{Hayden:2016cfa}
P.~Hayden, S.~Nezami, X.-L. Qi, N.~Thomas, M.~Walter, and Z.~Yang, {\it
  {Holographic duality from random tensor networks}},  {\em JHEP} {\bf 11}
  (2016) 009, [\href{http://arxiv.org/abs/1601.01694}{{\tt arXiv:1601.01694}}].

\bibitem{Bao:2018pvs}
N.~Bao, G.~Penington, J.~Sorce, and A.~C. Wall, {\it {Beyond Toy Models:
  Distilling Tensor Networks in Full AdS/CFT}},  {\em JHEP} {\bf 11} (2019)
  069, [\href{http://arxiv.org/abs/1812.01171}{{\tt arXiv:1812.01171}}].

\bibitem{Bao:2019fpq}
N.~Bao, G.~Penington, J.~Sorce, and A.~C. Wall, {\it {Holographic Tensor
  Networks in Full AdS/CFT}},  \href{http://arxiv.org/abs/1902.10157}{{\tt
  arXiv:1902.10157}}.

\bibitem{Akers:2018fow}
C.~Akers and P.~Rath, {\it {Holographic Renyi Entropy from Quantum Error
  Correction}},  {\em JHEP} {\bf 05} (2019) 052,
  [\href{http://arxiv.org/abs/1811.05171}{{\tt arXiv:1811.05171}}].

\bibitem{Dong:2018seb}
X.~Dong, D.~Harlow, and D.~Marolf, {\it {Flat entanglement spectra in
  fixed-area states of quantum gravity}},  {\em JHEP} {\bf 10} (2019) 240,
  [\href{http://arxiv.org/abs/1811.05382}{{\tt arXiv:1811.05382}}].

\bibitem{Harlow:2016vwg}
D.~Harlow, {\it {The Ryu\textendash{}Takayanagi Formula from Quantum Error
  Correction}},  {\em Commun. Math. Phys.} {\bf 354} (2017), no.~3 865--912,
  [\href{http://arxiv.org/abs/1607.03901}{{\tt arXiv:1607.03901}}].

\bibitem{Almheiri:2014lwa}
A.~Almheiri, X.~Dong, and D.~Harlow, {\it {Bulk Locality and Quantum Error
  Correction in AdS/CFT}},  {\em JHEP} {\bf 04} (2015) 163,
  [\href{http://arxiv.org/abs/1411.7041}{{\tt arXiv:1411.7041}}].

\end{thebibliography}\endgroup

\end{document}